\documentclass[12pt,a4paper]{article}
\usepackage{amsfonts}
\usepackage{amssymb}

\def\hybrid{\topmargin -20pt    \oddsidemargin 0pt
        \headheight 0pt \headsep 0pt
        \textwidth 6.25in       
        \textheight 9.5in       
        \marginparwidth .875in
        \parskip 5pt plus 1pt   \jot = 1.5ex}

\hybrid

\def\baselinestretch{1.2}

\catcode`\@=11

\def\marginnote#1{}
\newcount\hour
\newcount\minute
\newtoks\amorpm
\hour=\time\divide\hour by60
\minute=\time{\multiply\hour by60 \global\advance\minute by-\hour}
\edef\standardtime{{\ifnum\hour<12 \global\amorpm={am}%
        \else\global\amorpm={pm}\advance\hour by-12 \fi
        \ifnum\hour=0 \hour=12 \fi
        \number\hour:\ifnum\minute<10 0\fi\number\minute\the\amorpm}}
\edef\militarytime{\number\hour:\ifnum\minute<10 0\fi\number\minute}

\def\draftlabel#1{{\@bsphack\if@filesw {\let\thepage\relax
   \xdef\@gtempa{\write\@auxout{\string
      \newlabel{#1}{{\@currentlabel}{\thepage}}}}}\@gtempa
   \if@nobreak \ifvmode\nobreak\fi\fi\fi\@esphack}
        \gdef\@eqnlabel{#1}}
\def\@eqnlabel{}
\def\@vacuum{}
\def\draftmarginnote#1{\marginpar{\raggedright\scriptsize\tt#1}}

\def\draft{\oddsidemargin -.5truein
        \def\@oddfoot{\sl preliminary draft \hfil
        \rm\thepage\hfil\sl\today\quad\militarytime}
        \let\@evenfoot\@oddfoot \overfullrule 3pt
        \let\label=\draftlabel
        \let\marginnote=\draftmarginnote
   \def\@eqnnum{(\theequation)\rlap{\kern\marginparsep\tt\@eqnlabel}%
\global\let\@eqnlabel\@vacuum}  }

\def\preprint{\twocolumn\sloppy\flushbottom\parindent 2em
        \leftmargini 2em\leftmarginv .5em\leftmarginvi .5em
        \oddsidemargin -.5in    \evensidemargin -.5in
        \columnsep .4in \footheight 0pt
        \textwidth 10.in        \topmargin  -.4in
        \headheight 12pt \topskip .4in
        \textheight 6.9in \footskip 0pt
        \def\@oddhead{\thepage\hfil\addtocounter{page}{1}\thepage}
        \let\@evenhead\@oddhead \def\@oddfoot{} \def\@evenfoot{} }

\def\numberbysection{\@addtoreset{equation}{section}
        \def\theequation{\thesection.\arabic{equation}}}

\def\underline#1{\relax\ifmmode\@@underline#1\else
        $\@@underline{\hbox{#1}}$\relax\fi}

\def\titlepage{\@restonecolfalse\if@twocolumn\@restonecoltrue\onecolumn
     \else \newpage \fi \thispagestyle{empty}\c@page\z@
        \def\thefootnote{\fnsymbol{footnote}} }

\def\endtitlepage{\if@restonecol\twocolumn \else \newpage \fi
        \def\thefootnote{\arabic{footnote}}
        \setcounter{footnote}{0}}  

\catcode`@=12
\relax

\def\figcap{\section*{Figure Captions\markboth
        {FIGURECAPTIONS}{FIGURECAPTIONS}}\list
        {Figure \arabic{enumi}:\hfill}{\settowidth\labelwidth{Figure
999:}
        \leftmargin\labelwidth
        \advance\leftmargin\labelsep\usecounter{enumi}}}
 \relax
\def\tablecap{\section*{Table Captions\markboth
        {TABLECAPTIONS}{TABLECAPTIONS}}\list
        {Table \arabic{enumi}:\hfill}{\settowidth\labelwidth{Table
999:}
        \leftmargin\labelwidth
        \advance\leftmargin\labelsep\usecounter{enumi}}}
 \relax
\def\reflist{\section*{References\markboth
        {REFLIST}{REFLIST}}\list
        {[\arabic{enumi}]\hfill}{\settowidth\labelwidth{[999]}
        \leftmargin\labelwidth
        \advance\leftmargin\labelsep\usecounter{enumi}}}
 \relax
\makeatletter
\newcounter{pubctr}
\def\publist{\@ifnextchar[{\@publist}{\@@publist}}
\def\@publist[#1]{\list
        {[\arabic{pubctr}]\hfill}{\settowidth\labelwidth{[999]}
        \leftmargin\labelwidth
        \advance\leftmargin\labelsep
        \@nmbrlisttrue\def\@listctr{pubctr}
        \setcounter{pubctr}{#1}\addtocounter{pubctr}{-1}}}
\def\@@publist{\list
        {[\arabic{pubctr}]\hfill}{\settowidth\labelwidth{[999]}
        \leftmargin\labelwidth
        \advance\leftmargin\labelsep
        \@nmbrlisttrue\def\@listctr{pubctr}}}
 \relax
\makeatother
\newskip\humongous \humongous=0pt plus 1000pt minus 1000pt

\newif\ifdtup

\relax

\def\be{\begin{equation}}
\def\ee{\end{equation}}
\def\ba{\begin{eqnarray}}
\def\ea{\end{eqnarray}}

\def\del{\partial}

\def\r{\rho}
\def\a{\alpha}

\def\b{\beta}

\def\G{\Gamma}
\def\d{\delta}
\def\D{\Delta}

\def\m{\mu}
\def\n{\nu}
\def\om{\omega}

\def\l{\lambda}

\def\no{\noindent}

\def\qq{\qquad}

\def\IR{\relax{\rm I\kern-.18em R}}

\def \ha {{1\over 2}}

\def \ov {\over}

\def\IR{\relax{\rm I\kern-.18em R}}
\def\inv{^{\raise.15ex\hbox{${\scriptscriptstyle -}$}\kern-.05em 1}}

\def\tL{{\tilde L}}

\begin{document}

\renewcommand{\theequation}{\thesection.\arabic{equation}}

\newcommand{\beq}{\begin{equation}}
\newcommand{\eeq}[1]{\label{#1}\end{equation}}
\newcommand{\ber}{\begin{eqnarray}}
\newcommand{\eer}[1]{\label{#1}\end{eqnarray}}
\newcommand{\eqn}[1]{(\ref{#1})}
\begin{titlepage}
\begin{center}

\hfill CERN-TH/99-366\\
\vskip -.1 cm
\hfill NEIP-99-021\\
\vskip -.1 cm
\hfill hep--th/9912132\\
\vskip -.1 cm
\hfill December 1999\\

\vskip .3in

{\large \bf Domain walls of gauged supergravity, M-branes,\\ 
and algebraic curves}

\vskip 0.4in

{\bf I. Bakas${}^1$},\phantom{x} {\bf A. Brandhuber}${}^2$ \phantom{x}
and\phantom{x} {\bf K. Sfetsos${}^3$}
\vskip 0.1in
{\em ${}^1\!$Department of Physics, University of Patras \\
GR-26500 Patras, Greece\\
\footnotesize{\tt bakas@nxth04.cern.ch, ajax.physics.upatras.gr}}\\
\vskip .2in
{\em ${}^2\!$Theory Division, CERN\\
     CH-1211 Geneva 23, Switzerland\\
\footnotesize{\tt brandhu@mail.cern.ch}}\\
\vskip .2in
{\em ${}^3\!$Institut de Physique, Universit\'e de Neuch\^atel\\
Breguet 1, CH-2000 Neuch\^atel, Switzerland\\
\footnotesize{\tt sfetsos@mail.cern.ch}}\\

\end{center}

\vskip .4in

\centerline{\bf Abstract}

\no
We provide an algebraic classification of all supersymmetric 
domain wall solutions
of maximal gauged supergravity in four and seven dimensions, in the
presence of non-trivial scalar fields in the coset 
$SL(8,\IR)/SO(8)$ and $SL(5,\IR)/SO(5)$ respectively. 
These solutions satisfy first-order equations, which can be obtained 
using the method
of Bogomol'nyi. From  
an eleven-dimensional point of view 
they correspond to various continuous distributions of 
M2- and M5-branes. 
The Christoffel--Schwarz 
transformation and the uniformization of the associated algebraic
curves are used in order to determine the Schr\"odinger potential 
for the scalar and graviton fluctuations on the corresponding 
backgrounds. 
In many cases we explicitly solve the Schr\"odinger problem by employing 
techniques of supersymmetric quantum mechanics.
The analysis is parallel to the construction of domain walls 
of five-dimensional gauged supergravity, with  
scalar fields in the coset $SL(6,\IR)/SO(6)$, using algebraic curves
or continuous distributions of D3-branes in ten dimensions.
In seven dimensions, in particular, our classification of domain walls 
is complete for the full scalar sector of
gauged supergravity. 
We also discuss some
general aspects of D-dimensional gravity
coupled to scalar fields in the coset $SL(N,\IR)/SO(N)$.

\vfill
\end{titlepage}
\eject

\def\baselinestretch{1.2}
\baselineskip 16 pt
\noindent

\def\tT{{\tilde T}}
\def\tg{{\tilde g}}
\def\tL{{\tilde L}}


\section{Introduction}

Recent years have seen an increasing interest in gauged and
ungauged supergravities in various dimensions following the
conjectured duality between gauge theories and string/M-theory
\cite{Maldacena,Witten,GKP}.
The AdS/CFT correspondence offers the possibility to
understand strongly coupled gauge theories from a dual
supergravity description. The relevant backgrounds have
Poincar\'e invariance along the boundary directions and are
asymptotic to Anti-de Sitter (AdS) space. The deviations from
the AdS geometry in the interior correspond either to 
the broken phase of the theory due to non-zero vacuum expectation
values of scalar fields \cite{Maldacena}, \cite{MW}-\cite{CGLP}
or to deformations of the conformal field theory 
\cite{Akhme}-\cite{rey}.

In this paper we shall be concerned only with the first possibility.
In particular, we will construct a large class of solutions of $D=7$ and
$D=4$ supergravity that are dual to the 
$(2,0)$ theories in six dimensions \cite{Strominger}
and the three-dimensional theories with sixteen supersymmetries 
\cite{Seiberg} on the Coulomb branch.  
These theories correspond to the world volume theories of
parallel M5- and M2-branes respectively.
In contrast to the conformal cases, the solutions include also non-zero 
scalar fields, which, although they vanish at the boundary, become large 
in the interior. 
These solutions can be lifted to
eleven dimensions describing the gravitational field of distributions
of a large number of M2- or M5-branes. 
The location of the branes is directly related
to scalar Higgs expectation values on the field theory side.
Using such solutions, the spectrum of scalar and graviton excitations as 
well as the 
expectation values of Wilson loops can be calculated, shedding
new light on the AdS/CFT correspondence. 
Such investigations have been
carried out for supergravity duals of ${\cal N} = 4$ SYM, 
sometimes with surprising findings \cite{FGPW2,BS1,CR-GR}.
Interestingly, many of these backgrounds arise as limits of
charged AdS black holes in the lower-dimensional gauged 
supergravity theories or as extremal limits \cite{KLT,sfe1,Baksfe1} of
rotating brane solutions \cite{CVY2,KLT,RS1,CRST}; 
they have null singularities near the continuous distributions
in the higher dimensional backgrounds.
Such singularities arise generically in the flows from
conformal to non-conformal theories and it would 
be interesting to understand them better from the
field theory side.\footnote{Recently, 
supergravity duals of supersymmetric gauge theories with eight
supersymmetries were constructed and a mechanism for resolving
a certain type of space-time singularity
was proposed \cite{Polchinski}. However, 
the situation in \cite{Polchinski} is
different because, in ${\cal N} = 2$ SYM, the
classical superconformal point at the origin of the Coulomb branch
is removed by quantum effects, 
whereas the Coulomb branch of ${\cal N} = 4$ SYM
is uncorrected. Therefore, it is not expected that a similar
mechanism is at work for backgrounds with sixteen supersymmetries.}
Furthermore, these geometries can be viewed as 
examples of consistent truncations in various dimensions. The
embedding of lower-dimensional into higher-dimensional supergravities
has been worked out explicitly only for a couple of cases including
the $S^4$ and $S^7$ compactifications of 
eleven-dimensional supergravity \cite{CJ}
to $D=7$ \cite{Nastase} and $D=4$ \cite{WN1} gauged supergravity, 
respectively.
More examples of explicit Kaluza--Klein Ans\"atze 
which relate gauged supergravities to ten/eleven-dimensional
supergravity have been worked out in the recent papers \cite{Cveall}.

An additional motivation for investigating solutions of 
gauged supergravity in detail, as well as the spectra associated to 
quantum fluctuations, stems from the possibility to 
apply them to scenarios that view our world 
as a membrane embedded non-trivially in a higher dimensional 
non-factorizable space-time \cite{GW}.
This old idea has been revived recently in relation to the mass hierarchy 
problem \cite{rasu} (see also \cite{gobge}).
In the latter work a slice of the $AdS_5$
space, where our four dimensional world is embedded, was cut out,
thus resulting in a normalizable graviton zero mode. 
However, it turns out that there exists in addition a 
continuum of massive graviton modes with no mass gap separating them from 
the massless one. It was shown in \cite{rasu} that these modes have 
negligible effect to Newton's law. 
However, it is quite desirable in this context to find ways to
model possible modifications of Newton's law, 
since any deviations from it at the sub-millimeter scale have not been ruled 
out by the present day experiments (see, for instance, \cite{Long}).
It is known, on the other hand, that there are Yukawa-type 
modifications to Newton's law in theories that possess a 
mass gap separating massless from massive graviton modes. 
Besides the obvious phenomenological advantage of such models, 
the existence of a mass gap leads to a well-defined effective
field theory of the standard model of particles plus the massless graviton. 
This mass gap should be independent of the details of the slicing or 
at least be practically insensitive to them. 
Toy models possessing such desired features have already been 
constructed in \cite{BS2} 
in the context of five-dimensional gauged supergravity.
In this sense, many of the models, that we will describe 
in the present paper, as well as many of the models in \cite{Baksfe1}, 
can also be used to further pursue these ideas.

This paper is organized as follows: In section 2 we describe the bosonic 
sector of gauged supergravity
in $D$ dimensions. The non-zero scalars take values in the coset
space $SL(N,\IR)/SO(N)$ common to gauged supergravities in any
dimension. We restrict to vacua with $(D-1)$-dimensional Poincar\'e
invariance and find that for certain values of $D$ and $N$
the equations can be cast in first-order form. These Bogomol'nyi-type 
equations are equivalent to Killing spinor equations and
these solutions preserve sixteen supersymmetries. In section 3
we present solutions in arbitrary dimensions
that preserve part of the $SO(N)$ isometries and examine some of their 
general features. We explain how these solutions, for the cases that 
correspond to gauged supergravity, can be lifted
to string/M-theory, thus showing that they are consistent truncations
of ten- or eleven-dimensional supergravity respectively. These 
higher-dimensional 
backgrounds arise in the AdS/CFT correspondence as
supergravity duals of the field theories living on D3, M2 or M5 branes
on the Coulomb branch. 
In section 4 we give a general discussion of the scalar
field equations and graviton fluctuations in these backgrounds, and
the mass spectrum of operators in the dual field theory. 
Our discussion is facilitated by the fact that the corresponding equations 
can be cast into one-dimensional 
Schr\"odinger equations with appropriately chosen
potentials.
We also make contact with the theory of supersymmetric quantum mechanics
and outline the necessary elements that will be used for solving  
explicitly the Schr\"odinger problem in some cases of current interest.
Section 5 is devoted to a detailed discussion of the differential
equation for the conformal factor of the background metrics written in
conformally flat form.
We present a complete classification of solutions with different
unbroken isometries in terms of irreducible algebraic curves. The problem
of solving the domain-wall equations is then essentially reduced
to the uniformization of the relevant curves and the inversion of the 
corrsponding functions.
In Section 6 we treat in detail distributions of M2-branes, which
correspond to curves with genus $g \leq 1$. We present their uniformization
and obtain explicit expressions for the conformal factor
and the Schr\"odinger potential of the equation
for the scalar and graviton field fluctuations, whenever this is
possible in analytic form. 
In almost all cases of genus zero, the spectrum can be found exactly, whereas
for the other models we
utilize the WKB approximation to determine the mass gap and spacing
in the spectra. In section 7 we repeat the same analysis for distributions
of M5-branes. We also include, as application, the calculation of 
vacuum expectation values of Wilson surface operators in the six-dimensional
$(0, 2)$ theories on the Coulomb branch using the eleven-dimensional 
backgrounds.
Section 8 contains a summary of the Lam\'e equation
and its generalizations, which arise in the study of scalar and
graviton fluctuations in the background of domain walls associated
to elliptic functions. Elements of supersymmetric quantum mechanics 
are also used to expose connections between different elliptic
potentials, and discuss briefly some features of their exact 
spectrum beyond the WKB approximation.
Finally, we end with a short discussion in section 9 and
list some open problems.

The present paper generalizes previous work by two of the authors 
\cite{Baksfe1}, where all domain-wall solutions of 
five-dimensional gauged supergravity with non-trivial scalar fields 
in the coset $SL(6, \IR)/SO(6)$ were classified in terms of 
algebraic curves. In various places we include for completeness,
but with no further explanation, results on the $D=5$, $N=6$ 
theory, which correspond to continuous distributions of D3-branes
in ten-dimensional type-IIB supergravity.

\section{Gauged supergravity and first-order equations}
\setcounter{equation}{0}

We consider the sector of gauged supergravity theories with
non-trivial fields in the coset space $SL(N,\IR)/SO(N)$.
This is in general only a subset of a larger coset space and
is common to gauged supergravity theories in any dimension.
To be specific we will study the following Lagrangian in
$D$ dimensions
\be
{\cal L} = \frac{1}{4} {\cal R} - \frac{1}{2} \sum_{I=1}^{N-1} 
(\partial \a_I)^2 - P(\a_I)\ ,
\label{lagr}
\ee
although we will be mainly interested in the two cases
$(D,N) = (4,8)$ and $(D,N) = (7,5)$
that correspond to a sector of the 
four- and seven-dimensional gauged supergravities \cite{WN1,Nastase}.
Another case of interest, corresponding to a sector of
the five-dimensional gauged supergravity \cite{PPN,GRW}, is 
$(D,N) = (5,6)$.
All other fields including the
fermions are zero. In this subsector the scalar potential $P$ can
be expressed in terms of a symmetric $N\times N$ matrix $M = S^T S$, 
where $S$ is an element of $SL(N,\IR)$. Using an $SO(N)$
rotation matrix, $M$ can be diagonalized 
\be
M = {\rm diag} \{e^{2 \b_1}, \ldots, e^{2 \b_N} \} \ ,\qq
\sum_{i=1}^N \b_i=0\ .
\label{mbeta}
\ee
In addition, since the determinant of $M$ is 1, it 
depends on $N-1$ independent fields. 
We parametrize the fields $\b_i$ by the $N-1$ independent scalars
$\a_I$ in the following way
\be
\b_i = \sum_{I=1}^{N-1} \l_{iI} \a_I\ ,
\label{betamalpha}
\ee
where $\l_{iI}$ is an $N \times (N-1)$ matrix. The rows of this
matrix correspond to $N$ weights of the fundamental representation
of $SL(N)$, which obey the following normalizations:
\be 
\sum_{I=1}^{N-1} \l_{iI} \l_{jI} = 2 \delta_{ij} - \frac{2}{N}\ ,\qq
\sum_{i=1}^{N} \l_{iI} \l_{iJ} = 2 \delta_{IJ} \ ,\qq
\sum_{i=1}^N \l_{iI} = 0\ .
\label{weightsnorm}
\ee
Then the potential takes the form
\be
P= -\frac{g^2}{32} \left[ ({\rm tr} M)^2 - 2 {\rm tr} (M^2)
\right]\ .
\label{potentialm}
\ee

The equations of motion that follow from varying the action (\ref{lagr})
with respect to the metric $G_{MN}$ and the scalars $\a_I$ are
\ba
&&\frac{1}{4} {\cal R}_{MN} - \frac{1}{2} \sum_{I=1}^{N-1} \partial_M \a_I
\partial_N \a_I - \frac{1}{D-2} G_{MN} P = 0 \ , \nonumber \\
&&\partial_M ( \sqrt{-G} G^{MN} \partial_N \a_I)  -  \sqrt{-G} 
\frac{\partial P}{\partial \a_I} = 0\ .
\label{eom}
\ea
For the applications we have in mind we need the metric to exhibit 
$(D-1)$-dimensional Poincar\'e invariance, namely 
\ba
ds^2 & = & e^{2 A(z)} \left( \eta_{\m\n} dx^\m dx^\n + dz^2 \right) \ =
\nonumber\\
& = & e^{2 A(r)} \eta_{\m\n} dx^\m dx^\n + dr^2   \ ,
\label{metric}
\ea
where the ralation between the coordinates $z$ and $r$ is 
such that $dr=-e^A dz$. 
Furthermore we assume that the scalars depend only on $z$ and we
concentrate on solutions arising from first order equations.

The traditional way would be to consider
the Killing spinor first-order 
equations in the three different cases, when \eqn{lagr}
corresponds to a sector of gauged supergravity, 
as it was done for $(D,N)=(5,6)$ in 
\cite{FGPW2}.
That would leave half of the supersymmetries unbroken. 
However, there is an alternative way to
arrive at first-order field equations using a method \`a la 
Bogomol'nyi 
\cite{Cvetic}-\cite{Skenderis} that will be 
useful for further generalizations. 
For this task we have to plug the ansatz (\ref{metric})
into the action and rewrite the potential in terms of the
{\it prepotential} 
\be
W = -\frac{1}{4}{\rm tr} M = -{1\ov 4} \sum_{i=1}^N e^{2 \b_i} \ .
\label{prepotential}
\ee
We find
\be
P= \frac{g^2}{8} \left( \sum_{I=1}^{N-1} \left( 
\frac{\partial W}{\partial \a_I} \right)^2 - 4\left(1- {2 \ov N}\right)
W^2 \right)\ .
\label{potentialw}
\ee
The resulting one-dimensional effective action 
\be
S = \int dr e^{(D-1) A} \left( \sum_I \left( {d\a_I \ov dr} \right)^2 - 
\frac{1}{2} (D-1)(D-2)
\left( {dA \ov dr} \right)^2 + 2 P \right) \ ,
\label{effact}
\ee
can be written as a functional, which is a complete square plus a
boundary term
\ba
&& S  = \int dr e^{(D-1) A} \left(\sum_I \left( {d\a_I \ov dr} - \frac{g}{2} 
\partial_I W \right)^2 - {(D-1)(D-2) \ov 2} 
\left( {dA \ov dr} + \frac{g}{(D-2)} W \right)^2 \right) + \nonumber  \\
& & \phantom{xxxx}
+ \left. 2 g (D-2) e^{(D-1) A} W \right|_{r=-\infty}^\infty 
\label{bogo}\ ,
\ea
provided that the number of scalar fields and the dimension of space-time
are related as
\be
N = 4 {D-2\ov D-3}\ . 
\label{NNDD}
\ee
This relation will be used to simplify various expressions in the rest 
of the paper. 

Then, from \eqn{bogo} we can read off the first-order differential equations:
\be
\frac{dA}{dr}  =  -\frac{g}{D-2} W \ , \qq
\frac{d\a_I}{dr}  =  \frac{g}{2} \frac{\partial W}{\partial \a_I} \ .
\label{firstorder}
\ee
Note that \eqn{NNDD} has integer solutions only for 
the values of $N$ and $D$ 
that were mentioned 
at the beginning of this section, namely 
$(D,N) = (5,6)$, $(D,N) = (7,5)$ and $(D,N) = (4,8)$. In these three cases 
there exists a maximally supersymmetric solution of the equations of motion 
\eqn{eom} that preserves all 32 supercharges, in which all scalar fields are
set to zero and the metric is just $AdS_D$. 
In these cases the potential in 
\eqn{potentialm} becomes $P=-g^2 {N(N-2)/32}$ and equals by definition to the
negative cosmological constant. This defines the mass scale $g$ that we
have already used. The associated length scale $R$ defined by $g=2/R$
will also be used in this paper.
The same $AdS_D$ space is obtained for general values 
of $D$ and $N$ when all scalars 
are set to zero, but then, there is no notion of supersymmetry.

\section{The general solution}
\setcounter{equation}{0}

We begin this section with the construction of the most general solution of 
the non-linear system of equations within the $D$-dimensional ansatz 
\eqn{metric}
that preserves Poincar\'e invariance in the embedded 
$(D-1)$-dimensional space-time.
To keep our considerations as general as possible we will assume that the 
scalar field potential is given by \eqn{pottiw},
but we will not assume from the very beginning 
that the relation between the number of scalars and the dimensionality of the
space-time \eqn{NNDD} holds. 
After discussing some general properties of the corresponding 
configurations, we concentrate on those cases where 
the relation \eqn{NNDD} holds, and therefore the solutions originate  
from gauged supergravity in the appropriate number of dimensions. 
Then, we lift our solutions 
to eleven or ten dimensions in the context of eleven- or ten-dimensional 
supergravity. 
Our analysis is based on an analogous treatment of the case $(D,N)=(5,6)$  
that was developed in reference \cite{Baksfe1}.

\subsection{Solutions in arbitrary dimensions}

It is possible to find the most general solution of the coupled 
system of equations \eqn{firstorder}. 
As we have already mentioned, we will consider the potential arising 
from expression \eqn{potentialw} after using \eqn{NNDD}. We find that 
\be
P= \frac{g^2}{8} \left( \sum_{I=1}^{N-1} \left( 
\frac{\partial W}{\partial \a_I} \right)^2 - 2{D-1\ov D-2} W^2 \right)\ .
\label{pottiw}
\ee
Unless otherwise specified, we will no longer 
restrict $N$ and $D$ to obey \eqn{NNDD}. The resulting theory is that of $N-1$ 
scalars coupled to $D$-dimensional gravity with the interaction 
potential \eqn{pottiw}.
However, even though it no longer represents,
for arbitrary values of $N$ and $D$, a sector of some gauged
supergravity theory, it still admits interesting solutions arising from the 
first-order Bogomol'nyi equations \eqn{firstorder}.

In order to proceed further,
we first compute the evolution of the auxiliary scalar fields $\b_i$.
Using \eqn{betamalpha} and \eqn{firstorder} we find 
\be
\b_i^\prime = 2 {D-2\ov N}
 A^\prime + {g \ov 2}  e^{2 \b_i + A}\ ,
\qq i=1,2,\dots , N\ ,
\label{eh10}
\ee
where the prime denotes derivative with respect to the argument $z$.
It is easy to integrate these $N$ decoupled first-order equations for the
$\b_i$'s. For further convenience 
we reparametrize the function $A(z)$ in terms of an 
auxiliary function $F(z g^2)$ as follows
\be
e^A = g (-F^\prime)^{\frac{N}{4 (D-2)+N}} \ ,
\label{ai10}
\ee 
where the prime denotes here derivative with respect to the 
argument $z g^2$. The minus sign we have included in this definition
implies that, for consistency, the function $F$ should be monotonously 
decreasing with $z$.
Then, according to this ansatz we find that the general solution 
for the $D$-dimensional metric \eqn{metric} is 
\be
ds^2  =  {g^2 f^{1 \ov 2 (D-2)}} \eta_{\m\n}dx^\m dx^\n 
+ g^{-2} f^{-{2\ov N}} dF^2\ ,
\label{fh300}
\ee
with
\be 
f = \prod_{i=1}^N (F-b_i)\ ,
\label{hj3}
\ee
and the solution for the scalar fields in \eqn{eh10} 
is given by 
\be
e^{2 \b_i} = {f^{1/N}\ov F-b_i}\ ,\qq i=1,2,\dots , N\ .
\label{hja10}
\ee
The $b_i$'s are $N$ constants of integration,
which can be ordered as
\be
b_1\geq b_2\geq \dots \geq b_N\ ,
\label{hord10}
\ee
without loss of generality.
Also, since the sum of the $\b_i$'s
is zero, we find that 
the function $F$ has to satisfy the differential equation
\be
(-F^\prime)^\D =\prod_{i=1}^N (F-b_i)=f \ ,\qq \D={4 (D-2)N \ov 4(D-2) + N}\ .
\label{jds100}
\ee

If we assume that our models arise from gauged 
supergravity, and therefore the relation \eqn{NNDD}
between the number of scalar fields $N$ and
the dimensionality of space-time $D$ holds,
the metric \eqn{fh300} will become 
\be
ds^2  =  {g^2 f^{1 \ov 2 (D-2)}} \eta_{\m\n}dx^\m dx^\n 
+ g^{-2} f^{-{D-3\ov 2(D-2)}} dF^2\ ,
\label{fh30}
\ee
whereas the scalar fields are still given by \eqn{hja10}.
The function $F(z g^2)$ is determined by solving the equation
\be
(F^\prime)^4 =\prod_{i=1}^N (F-b_i)=f \ ,
\label{jds10}
\ee
since the constant $\D$ equals 4 in all three cases 
$(D, N) = (4, 8)$, $(5, 6)$ and $(7, 5)$.
Although the relation \eqn{NNDD} guarantees  
that our models originate from gauged supergravity, 
it should be emphasized that it is not the 
only possibility that results into an integer value for $\D$. 
As an example, choosing $D=5$ and $N=4$ we obtain $\D=3$. Another
interesting case, but with non-integer $\D$, arises for $D=3$ 
and $N=2$: it has $\D = 4/3$ and describes three-dimensional 
gravity coupled to a single scalar 
in the presense of a negative cosmological constant (since the
potential \eqn{pottiw} turns out to be constant).

So far, we have presented our general solution in a coordinate 
system where $F$ is viewed as the independent variable.\footnote{To 
compare this with the results of \cite{Baksfe1} for the values 
$D=5$ and $N=6$, we should replace the function $F(z/R^2)$ and the 
constants $b_i$ used in \cite{Baksfe1} by
$ 4 F(z g^2 )$ and $4 b_i$, respectively.}
If we insist on presenting the solution in a conformally flat form,  
as given by the first line in \eqn{metric}, 
the differential equation \eqn{jds100} needs to be solved 
to obtain $F(z g^2)$. 
This will be studied in detail in section 5, using the theory of 
algebraic curves and their uniformization, as it is a necessary 
step for investigating the Schr\"odinger equations that arise for 
the massless scalar and graviton fluctuations.

If the constants $b_i$ are all equal, our solution becomes nothing 
but $AdS_D$ (with radius $2 (D-2) R/N$) 
with all scalar fields turned off to zero.
In the opposite case, 
when all constants $b_i$ are different from one another, there is
no continuous subgroup of $SO(N)$ preserved by our solution.
If we let some of the $b_i$'s coincide, we restore various 
continuous subgroups of $SO(N)$ accordingly.
Imposing the reality condition on the scalars in \eqn{hja10}
restricts the values of $F$ to be larger that the maximum of the constants 
$b_i$; then, according to the ordering in \eqn{hord10}, 
this means that $F\ge b_1$.
For $F\gg b_1$ the scalars tend to zero and $f\simeq F^N$, in which case 
the metric in \eqn{metric} approaches $AdS_D$ (with radius $2  (D-2) R/N$) 
as expected. In the conformally flat form of the metric, where $z$ is  
viewed as the independent variable, we have $e^A\sim 1/z$ and therefore 
$z=0$ is taken as the origin of the $z$-coordinate.
Hence,  
in solving the differential equation \eqn{jds10}, we will choose
the constant of integration  
so that in the limit $F\gg b_1$ we have the asymptotic behaviour
$F\sim 1/z^{4(D-2)/N}$.\footnote{Using \eqn{hja10} we can show that
the scalar 
fields go to zero as $\a_I\sim z^{D-3}$, when \eqn{NNDD} is obeyed.
This is consistent with the interpretation that these scalars 
parametrize states in the Coulomb branch of the $N=4$ SYM theory and they
do not correspond to explicit mass deformations.}
For intermediate values of $F$ we have a flow in the $D$-dimensional 
space spanned by all scalar fields $\b_i$. 
In general we may have $b_1=b_2=\dots = b_n$, 
with $n\leq N$, when $b_1$ is $n$-fold degenerate.
In this case, the solution preserves an $SO(n)$ subgroup of $SO(N)$ 
and the flow is actually taking place in $N-n$ dimensions. 
On the other hand, let us consider
the case when $F$ approaches its lower value $b_1$.
Then, the scalars in \eqn{hja10} are approaching 
\ba
e^{2\b_i}\simeq  \left \{ \begin{array} {ccc}
 f_0^{1/N} (F-b_1)^{(n-N)/N}\ , \quad & {\rm for} \ \ & i=1,2,\dots , n \\ 
\\
{f_0^{1/N}\ov b_1-b_i} (F-b_1)^{n/N}\ ,\quad & {\rm for} \  \ & 
i=n+1,\dots , N \\
\end{array} \right \} \ ,
\label{jkef1}
\ea
where $f_0=\prod_{i=n+1}^N (b_1-b_i)$. Consequently, we have a 
one-dimensional flow in this limit since the scalar fields $\b_i$ can be 
expressed in terms of a single (canonically normalized) scalar $\a$ as 
\ba
{\bf \vec{\b}}  & \simeq & \sqrt{2\ov N n(N-n)} 
(n-N,\dots, n-N,n,\dots,n)\ \a\ ,
\nonumber\\
\a & \simeq  &\ha \sqrt{n(N-n)\ov 2 N} \ln(F-b_1)\ .
\label{fjk23}
\ea

When our solutions correspond to gauged supergravities and the
relation \eqn{NNDD} applies, it is 
also useful to find the limiting form of the metric \eqn{fh30} when
$F\to b_1$. Changing the variable to $\r$ as
\be
F= b_1 + \left({(1-n/N) f_0^{1/N} g\r}\right)^{N\ov N-n} \ ,
\label{jdr23}
\ee
the metric \eqn{fh30} becomes, for $\r\to 0^+$ 
\be
ds^2 \simeq d\r^2 + \left((1-n/N)^n f_0 g^{(D-3)(N-n)+n}
\right)^{2\ov (D-3)(N-n)}
\r^{2 n\ov (D-3)(N-n)} \eta_{\m\n} dx^\m dx^\n \ .
\label{jksd1}
\ee
Hence, at $\r=0$ (or equivalently at $F=b_1$) there is a naked singularity
that can be interpreted, as we will see later in a higher-dimensional 
context, as the location of a distribution of M5-, M2- or 
D3-branes for $(D, N)=(7,5)$, $(4,8)$ and $(5,6)$ respectively. 
It will also be seen that 
this singularity is time-like for $n=1,2,3$ and
null for $4\leq n<N$.
%

\subsection{M-theory branes}

It is possible to lift our seven- and four-dimensional solutions
with metric and scalars given by \eqn{fh30} and \eqn{hja10}
with $(D,N)=(7,5)$ and $(4,8)$, respectively,
to supersymmetric solutions of eleven-dimensional supergravity, 
where only the metric and the three-form are turned on. 
The eleven-dimensional solutions will correspond to the 
gravitational field of a large number of M5-branes and M2-branes 
in the field theory limit
with a special continuous 
distribution of branes in the transverse to the brane space.
It shows that they are true compactifications of 
eleven-dimensional supergravity on $S^4$ or $S^7$, respectively; 
this is also what is expected on general grounds \cite{witnik}.
The method we follow is the same as that used in \cite{Baksfe1}
to lift the 
five-dimensional solution with $(D,N)=(5,6)$ to a supersymmetric solution of
the ten-dimensional type-IIB supergravity representing distributions of 
D3-branes in the field theory limit. A brief summary of the main results
will also be included for 
completeness in this case. 

The higher-dimensional 
metrics for the various distributions of branes have the form
\be
{\rm M5\!-\! brane}:\qq 
ds^2 = H_0^{-1/3} \eta_{\m\n} dx^\m dx^\n + H_0^{2/3} (dy_1^2+dy_2^2+\dots +
dy_5^2)\ ,
\label{M55}
\ee
\be
{\rm M2\!-\! brane}:\qq 
ds^2 = H_0^{-2/3} \eta_{\m\n} dx^\m dx^\n + H_0^{1/3} (dy_1^2+dy_2^2+\dots +
dy_8^2)\ ,
\label{M22}
\ee
and 
\be
{\rm D3\!-\! brane}:\qq 
ds^2 = H_0^{-1/2} \eta_{\m\n} dx^\m dx^\n + H_0^{1/2} (dy_1^2+dy_2^2+\dots +
dy_6^2)\ .
\label{D33}
\ee
In all cases $H_0$ is a harmonic function 
in the $N$-dimensional space $\IR^N$
transverse to the brane parametrized by the $y_i$ coordinates.
However, instead of being asymptotically flat, the metrics \eqn{M55}--\eqn{D33}
will become asymptotically $AdS_D\times S^{N-1}$ for large radial distances,
with $D$ and $N$ taking their appropriate values. The radius of the sphere is 
always $R$, whereas, in agreement with our previous normalization, it is  
$\ha(D-3) R$ for $AdS_D$.
Under these conditions, the higher-dimensional solutions break half of the 
maximum  of 32 supersymmetries (see, for instance, \cite{KaKou}).
It is only for coinciding branes, when the metric is exactly 
$AdS_D\times S^{N-1}$, that the maximal number of supersymmetries is preserved 
and the backgrounds are presumably exact vacua.

We proceed further by first performing the coordinate change 
\be
y_i = 2 g^{(D-5)/2} (F-b_i)^{1/2} \hat x_i\ ,\qq i=1,2,\dots , N\ ,
\label{ejh1}
\ee
where the $\hat x_i$'s define a unit $N$-sphere.
It can be shown that the $N$-dimensional flat metric
in the transverse part of the brane metric \eqn{M55}--\eqn{D33} 
can be written as
\be
 \sum_{i=1}^N dy_i^2 = g^{D-5} \sum_{i=1}^N {\hat x_i^2\ov F-b_i}\ dF^2
+4 g^{D-5} \sum_{i=1}^N (F-b_i) d\hat x_i^2\ .
\label{dm22}
\ee
The harmonic function $H_0$ is determined by 
comparing the massless scalar equation 
for the eleven- and ten-dimensional metrics \eqn{M55}--\eqn{D33}, 
with the same equation arising 
using the D-dimensional metric \eqn{fh30}.
In both cases one makes the ansatz that the solution 
does not depend on the sphere coordinates, i.e. $\Phi= e^{i k \cdot x}
\phi(z)$.
Since the solutions for the scalar $\Phi$ should be the same in 
any consistent truncation of the theory, 
the resulting second-order ordinary differential
equations should be identical. 
A comparison allows to determine the function $H_0$ as follows
\be
H_0^{-1}  = {4 \ov R^4}  f^{1/2} \sum_{i=1}^N {y_i^2\ov (F-b_i)^2}\ .
\label{dhj1}
\ee
The coordinate $F$ is determined in terms of the transverse coordinates
$y_i$ as a solution of the algebraic equation
\be
\sum_{i=1}^N  {y_i^2\ov F-b_i} =4 g^{D-5}\ .
\label{jk4}
\ee
Note that the algebraic equation \eqn{jk4} for $F$ cannot be solved 
analytically for general choices of the constants $b_i$.
However, this becomes possible when some of the $b_i$'s coincide
in such a way that the
degree of the corresponding algebraic equations is reduced to 4 or less. 
We also note that $H_0$ as defined in \eqn{dhj1} and \eqn{jk4}
is indeed a harmonic function in $\IR^N$. The proof was made in \cite{Baksfe1}
for the case $N=6$, but even if $N$ is kept general all intermediate steps 
are essentially the same. 
We note that one may use as independent variables, instead of the $y_i$'s,
spherical coordinates since the constraint
\eqn{jk4} is then automatically satisfied.
For D3-branes, this was done for the various cases of interest  
in \cite{Baksfe1}. For M2- and M5-branes 
we expect to recover the metrics found in \cite{CGLP}.
Various subcases were also 
considered before in \cite{KLT,sfe1,FGPW2,BS1} in connection with the 
Coulomb branch of gauge theories at strong coupling.

Brane solutions that are asymptotically flat
are obtained by replacing $H_0$ in \eqn{M55}--\eqn{D33} by $H=1+H_0$. 
Then, in this context, the length parameter 
$R$ has a microscopic interpretation using 
the eleven-dimensional Planck scale $l_{\rm P}$ or the string scale 
$\sqrt{\a'}$ and the string coupling $g_s$, 
and the (large) number of branes $N_{\rm b}$. For M5-branes we have
$R^3 =  \pi N_{\rm b} l_{\rm P}^3$, for M2-branes 
$R^6=32 \pi N_{\rm b} l_{\rm P}^6$, and for D3-branes 
$R^4= 4\pi N_{\rm b} g_s \a'^2$.

\section{The spectrum of fluctuations}
\setcounter{equation}{0}

In this section we investigate the problem of solving the differential 
equations that arise for the massless scalar field as well as for the 
graviton fluctuations in our 
general D-dimensional background metrics \eqn{fh30}.
After presenting some general features, we make contact with 
supersymmetric quantum mechanics, which will prove useful for making further 
progress in computing the exact spectrum later in sections 6, 7 and 8.

\subsection{Generalities}

Within the AdS/CFT correspondence \cite{Maldacena,Witten,GKP}
(assuming \eqn{NNDD}), 
the solutions and eigenvalues 
of the massless scalar equation
have been associated, on the gauge theory side, 
with the spectrum of the operator
${\rm Tr} F^2$, whereas those of the graviton fluctuations 
polarized in the directions parallel to the brane,
with the energy momentum tensor $T_{\m\n}$ \cite{GKT,Witten,GKP,FFZ}.
A priori these two spectra are different. However, it can be shown, 
similarly to the case of five-dimensional gauged supergravity investigated 
in \cite{BS2}, that
the two spectra and the corresponding eigenfunctions coincide.
Hence, in what follows, $\Phi$ will denote either a massless
scalar field or any component of the graviton tensor field.

We proceed further by making the following ansatz for the solution
\be 
\Phi(x,z) =\exp(ik\cdot x) \exp\left(-{D-2\ov 2} A\right) \Psi(z)\ ,
\label{hd22}
\ee
which represents 
plane waves propagating along the $(D-2)$-brane 
with an amplitude function that is $z$-dependent.
The mass-square is defined as $M^2 = - k\cdot k$. 
Then the equation for $\Psi$ can be cast into a time-independent 
Schr\"odinger equation for a wave function
$\Psi(z)$ as 
\be
-\Psi^{\prime\prime} + V \Psi = M^2 \Psi \ ,
\label{ss2}
\ee
with a potential given by 
\be
V  =  {(D-2)^2 \ov 4} {A^{\prime}}^2 + {D-2\ov 2} A^{\prime\prime}\ .
\label{jwd}
\ee
It should be noted and will be further discussed later in this section,
that this potential is of the form that appears in 
supersymmetric quantum mechanics and therefore the spectrum is non-negative
\cite{SQMwit,SQMrev}.\footnote{In the context of 
five-dimensional gauged supergravity this has been first hinted
in \cite{Baksfe1} and explicitly noted in \cite{DeWolfe}.}
However, this does not guarantee that the massless mode is normalizable.
In our case, $\Psi_0 \sim \exp((D-2)A/2)$ and it is clearly not 
square-normalizable
due to its behaviour near $z=0$, namely $\Psi_0\sim 1/z^{(D-2)/2}$.
Using the general formulae \eqn{ai10} and \eqn{jds100} 
it turns out that the potential takes the form 
\ba 
V  =  {f^{2/\D}\ov 4 R^4} \left[\left(1+{8\ov \D}\right) 
\Big(\sum_{i=1}^{N} {1\ov F-b_i}\Big)^2 
-8 \sum_{i=1}^{N} {1\ov (F-b_i)^2}\right]\ .
\label{hf8}
\ea
We note that in deriving \eqn{hf8} from \eqn{jwd} the 
relation \eqn{NNDD} has not been used. 

The expression \eqn{hf8} for the potential depends, of course, on the variable 
$z$ through the function $F(zg^2)$. Even without any knowledge of 
the explicit $z$-dependence
of the potential, we may deduce some general properties about the spectrum 
in the various cases of interest. 
In general, $F$ takes values between the maximum 
of the constants $b_i$ (which according to the ordering made in \eqn{hord10}
is taken to be $b_1$) and $+\infty$. 
When $F\to +\infty$ (or equivalently $z\to 0^+$), the space approaches
$AdS_D$ and the potential becomes
\be
V\simeq {D(D-2)\ov 4}\ {1\ov z^2}\ , \qq {\rm as}\quad z\to 0^+\ ,
\label{hjf4}
\ee
and hence it is unbounded from above.
Let us now consider the behaviour of the potential close to the other end,
namely when $F\to b_1$. Consider the general case 
where $b_1$ appears $n$ times, as
in the corresponding discussion made at the end of subsection 3.1.
Using \eqn{hf8} we find that the potential 
behaves (including the subscript $n$ 
to distinguish among various cases) as
\be
V_n\simeq {f_0^{2/\D}\ov 4 R^4} \left((1+8/\D)n^2 -8n\right) 
(F-b_1)^{{2 n\ov \D}-2}\ ,\qq
{\rm as} \quad F\to b_1\ ,
\label{hjfg3}
\ee
with $f_0$ being a constant given, as before, by $f_0=\prod_i(b_1-b_i)$.
Hence, for $\D< n\leq N$, the potential goes 
to zero and the spectrum is continuous. For $n=\D$ the potential 
approaches a constant value, which is given by 
$V_{\D,{\rm min}}={1\ov 4 R^4}\D^2 f_0^{2/\D} $.
Therefore, although the spectrum is continuous,
it does not start from zero, but there is a mass gap whose squared 
value is given by the minimum of the potential.
Hence, for $\D\leq n\leq N$ the range of $z$ necessarily extends to $+\infty$,
i.e. $0\leq z < \infty$, with $z=\infty$ corresponding to a null naked 
singularity (except in the AdS case where $n=N$).
If $\Delta<n<N$ we have that $(F-b_1)^{1-n/\D}\simeq
(n/\D-1) g^2 f_0^{1/\D} z$ near $F=b_1$, and therefore the potential
behaves as
\ba
\D<n<N: \qq 
V_n &\simeq&  {C_{n,\D}\ov z^2}\ ,
\qq {\rm as} \quad z\to \infty\ ,
\nonumber\\
C_{n,\D}&=&{1\ov 64} \left(n(\D+4)-4 \D\ov \D-n\right)^2 -{1\ov 4}\ .
\label{hjf24}
\ea
For $n=\D$ we have instead that $F-b_1\simeq e^{-f_0^{1/n} g^2 z}$, as 
$z\to \infty$ with the potential reaching the constant value that we have
mentioned above.
For $n< \D$ the potential goes to either $+ \infty$ or $-\infty$,
as $F\to b_1$
and therefore the spectrum cannot be continuous but discrete.
Therefore there should be a 
maximum value for $z$, denoted by $z_{\rm max}$, that is
determined by solving 
the algebraic equation $F(z_{\rm max}g^2)=b_1$. The value of $z=z_{\rm max}$
corresponds to a time-like naked singularity.
Then, using the relation $(F-b_1)^{1-n/\D}\simeq
(1-n/\D) g^2 f_0^{1/\D} (z_{\rm max}-z)$ near $F=b_1$, we find that 
\ba
n<\D: \qq 
V_n &\simeq&  {C_{n,\D}\ov (z-z_{\rm max})^2}\ ,
\qq {\rm as} \quad z\to z^-_{\rm max}\ ,
\nonumber\\
C_{n,\D}&=&{1\ov 64} \left(n(\D+4)-4 \D\ov \D-n\right)^2 -{1\ov 4}\ .
\label{hjf23}
\ea
One may worry that there are cases where the potential goes to $-\infty$ at 
$z=z_{\rm max}$ with a coefficient $C_{n,\D}$
that is smaller than $-1/4$; then, it is well known from elementary 
quantum mechanics, that the 
spectrum might be unbounded from below. 
However, this does not happen because 
we are dealing with supersymmetric quantum mechanics and the 
spectrum is bounded from below by zero. 
In addition, the coefficient $C_{n,\D}$ in
\eqn{hjf23} is manifestly greater or equal than $-1/4$; 
the limiting value $-1/4$ is reached for $n=4\D/(\D+4)$.
The qualitative analysis of the spectrum we just presented agrees with 
that performed for $(D,N)=(5,6)$ in \cite{Baksfe1}
and for $(D,N)=(7,5)$ and $(D,N)=(4,8)$ in \cite{CGLP}.

As we see later, solving the Schr\"odinger equation \eqn{ss2} 
and determining
the spectrum is a non-trivial problem, except for a few particular cases
where explicit calculations can be carried out in detail.
However, when the spectrum is discrete, as for $n<\D$,
we may use the approximate method of WKB. 
Since the quantum mechanical potentials are supersymmetric,
WKB is expected to be an excellent approximation, 
not only for high quantum 
numbers, but also for low ones \cite{SQMrev}. Moreover, if the potential 
turns out to be shape-invariant, the WKB approximation
is also exact \cite{SQMrev}.
It is convenient at this point to 
use $F$ as an independent variable instead of $z$. 
Then, changing the dependent variable 
in \eqn{ss2} as $\Psi=e^{(D-2)A/2}\phi$, we find the equation
\be
g^4 \del_F f^{{1\ov 4} +{1\ov \D}} \del_F \phi + M^2 f^{{1\ov 4}-{1\ov \D}}
\phi = 0\ .
\label{kasjf1}
\ee
Using well-developed methods for studying this type of differential 
equations (see, for instance, \cite{RS1}), 
the spectrum is found to be given approximately by
\be
M^2_m = {\pi^2\ov z_{\rm max}^2} m \left(m+ {D-3\ov 2} + {|(n/4-1)\D + n|
\ov 2(\D-n)}\right) + {\cal O}(m^0)\ ,\qq m=1,2,\dots \ ,
\label{jkw22}
\ee
where, as usual, $z_{\rm max} $ is the maximum value of $z$ and 
$\D$ is the constant defined in \eqn{jds100}.
It turns out that the validity of the WKB approximation 
requires that the inequality $n<\D$ be satisfied. This is also 
consistent with the fact that the potential should be unbounded at the 
end points $F=+\infty$ and $F=b_1$.

\subsection{Relation to supersymmetric quantum mechanics}

We have already mentioned that the potential \eqn{jwd} has the form  
encountered in supersymmetric quantum mechanics 
\cite{SQMwit,SQMrev}. Let us make this relationship more explicit by first
recalling that in supersymmetric quantum mechanics, two potentials are
supersymmetric partners of one another 
provided that there is a superpotential
$W(z)$ so that 
\be
V_1(z) = W^2 - W^{\prime} ~, \qq  V_2(z) = W^2 + W^{\prime} ~. 
\label{v1v2}
\ee
Then, in terms of the ladder operators
\be
\label{aad2}
a = {d \over dz} + W(z) ~, \qq a^{\dagger} =-{d \over dz} +W(z) \ ,
\ee
the eigenstates of the two Hamiltonians $H_1$, $H_2$ 
are related to each other as
\be
\psi_n^{(2)} = {1 \over \sqrt{E_{n+1}^{(1)}}} a \psi_{n+1}^{(1)} ~, \qq
 \psi_{n+1}^{(1)} = {1 \over \sqrt{E_n^{(2)}}} a^{\dagger} 
\psi_n^{(2)} ~, 
\label{hdh4}
\ee
whereas for the energy levels, in the case that the spectra of the two partner 
potentials are discrete, we have the relation
\be
E_n^{(2)} = E_{n+1}^{(1)} ~,\qq n = 0, 1, 2, \dots \ ,
\ee
with $E_0^{(1)} = 0$. Notice that if $\psi_{n+1}^{(1)}$ of $H_1$ is
normalized then the wave-function $\psi_{n}^{(2)}$ of $H_2$ will be
also normalized and vice-versa.  
Thus, on general grounds, knowing the spectrum of $H_1$ one can 
construct the spectrum of $H_2$; conversely, from the spectrum of 
$H_2$ one can deduce the spectrum of $H_1$ apart from the ground
state with $E_0^{(1)}=0$ which is not paired. This relation is true
only for the case of unbroken supersymmetry. When supersymmetry is 
broken, there is a 1-1 pairing of all eigenstates of $H_1$ and $H_2$
and the relations become modified by replacing $\psi_{n+1}^{(1)}$ 
with $\psi_n^{(1)}$ and $E_{n+1}^{(1)}$ with $E_n^{(1)}$ in the 
equations above. Then, the
potentials $V_1$ and $V_2$ have degenerate positive ground state
energies. 
We also recall that when two partner potentials have
continuum spectra the corresponding reflection and transmission 
probabilities are identical.

The Schr\"odinger potential \eqn{jwd} that arose in the study of quantum 
fluctuations on domain wall backgrounds has indeed the form $V_2(z)$ 
with superpotential 
\be
W(z) = {D-2 \over 2} A^{\prime}(z) \ .
\label{wdds}
\ee
Then, using \eqn{ai10} and \eqn{jds100} we find that the 
partner potential $V_1(z)$ in \eqn{v1v2} takes the form 
\ba 
V_1  =  {f^{2/\D}\ov 4 R^4} \left[\left(1-{8\ov \D}\right) 
\Big(\sum_{i=1}^{N} {1\ov F-b_i}\Big)^2 
+8 \sum_{i=1}^{N} {1\ov (F-b_i)^2}\right]\ .
\label{hf8s}
\ea
which is analogous to the form \eqn{hf8} for $V_2$. Specific examples
of this relation will be considered in detail in later sections.


\section{Algebraic classification}
\setcounter{equation}{0}

The underlying mathematical structure for solving the differential
equation 
\be
(F^{\prime}(z))^4 = (F(z) - b_1)(F(z) -b_2) \cdots (F(z) -b_N)\ ,
\ee
with arbitrary moduli $b_i$, is that of the Christoffel--Schwarz
transformation in complex analysis. This transformation is 
familiar from electrostatics, where one applies the technique
to find the electric potential for a given distribution of 
charges. From  
this point of view, it is not surprising that the
ansatz we made for constructing static domain walls in theories
of gauged supergravity amounts to solving a similar mathematical
problem. It is useful to think of the variable $z$ as being 
complex, whereas $F$ takes values in the complex upper-half 
plane. Of course, appropriate restrictions have to be made at 
the end in order to ensure the reality of the variable $z$ and
hence the reality of our domain wall solutions. As we will see in
detail, the solutions are characterized by the uniformization
of Riemann surfaces, which are naturally associated to the 
Christoffel--Schwarz transformation. Hence, the explicit
derivation of the Schr\"odinger potential $V(z)$ 
requires, for all practical purposes, going
through such a mathematical framework, apart from its own value
in providing a systematic classification of all domain-wall 
solutions in terms of algebraic curves. Also note that the 
variable $F$ is a function $F(zg^2)$, but for simplicity we set
$g=1$ in the following; this parameter can be easily reinstated
at the end by appropriate scaling in $z$. Here, we do not assume
any particular ordering of $b_i$.

We will treat the Christoffel--Schwarz transformation in a unified
way for all three cases of interest, namely $(D,N)=(4,8)$ (M2-branes),
$(D,N)=(5,6)$ (D3-branes), and $(D,N)=(7,5)$ (M5-branes), 
since there is a hierarchy of algebraic curves within this 
transformation that depends on the isometry groups
of the distributions of branes. It is useful to start with $N=8$
and consider an octagon in the complex $z$-plane, which is mapped
onto the upper-half plane via a Christoffel--Schwarz transformation
\be
{dz \over dF} = (F-b_1)^{-{\varphi}_1/\pi} 
(F-b_2)^{-{\varphi}_2/\pi} \cdots (F-b_8)^{-{\varphi}_8/\pi} ~.  
\ee
This transformation maps the vertices of the octagon 
to the points $b_1, b_2, \cdots , b_8$ on the real axis of the 
complex $F$-plane, whereas its interior 
is mapped onto the entire upper-half
$F$-plane. The variables ${\varphi}_i$ denote the exterior 
(deflection) angles of the octagon at the corresponding vertices,
which are constrained by geometry to satisfy the relation
${\varphi}_1 + {\varphi}_2 + \cdots + {\varphi}_8 = 2\pi$.
We proceed by making the canonical choice of angles
${\varphi}_1 = {\varphi}_2 = \cdots = {\varphi}_8 = \pi/4$, in 
which case we arrive at the differential equation that relates
$dz$ and $dF$: 
\be
\left({dz \over dF}\right)^4 = (F-b_1)^{-1}(F-b_2)^{-1} \cdots
(F-b_8)^{-1} ~, 
\ee
which is the equation we have to solve for the case of M2-branes. 

It is convenient at this point to introduce complex algebraic 
variables
\be
x=F(z) ~, ~~~~~ y =F^{\prime}(z) ~, 
\ee
which cast the above differential equation into the form of an 
algebraic curve 
\be
y^4 = (x-b_1)(x-b_2) \cdots (x-b_8) ~. 
\ee
This defines a Riemann surface of genus $g=9$,\footnote{We will use the 
symbol $g$ to denote the genus of a Riemann surface. The fact that 
the same notation has already been used for a mass scale should not create a
confusion to the reader.}
when any two moduli 
are not equal, as follows by direct application of the Riemann--Hurwitz
relation, which is standard in algebraic geometry.
The task now is to uniformize the algebraic curve by finding
another complex variable, call it $u$, so that $x = x(u)$ and 
$y = y(u)$, which resolves the problem of multi-valuedness of the 
algebraic equation above; the corresponding Riemann surface is 
pictured geometrically by gluing four sheets together along their 
branch cuts. 
Then, following the definition of $x$ and $y$ in terms of $F(z)$
and its $z$-derivative, one applies the chain rule in order to 
obtain the function $z(u)$ by integration of 
the resulting first-order equation
\be
{dz \over du} = {1 \over y(u)}{dx(u) \over du} ~. 
\label{ssool}
\ee
Finally, by inverting the result we obtain the function $u(z)$, 
which yields $F(z)$, and hence the conformal factor of the 
corresponding domain wall solutions, as well as the Schr\"odinger 
potential $V(z)$ for the graviton and scalar field fluctuations
in these backgrounds. Of course, there is an integration constant
that appears in the function $z(u)$, but this can be fixed by 
requiring that the asymptotic behaviour of the domain walls  
approach the $AdS$ geometry as $z \rightarrow 0$. We also note for
completeness that there is a discrete symmetry $x \leftrightarrow -x$,
$b_i \leftrightarrow -b_i$ that leaves invariant the form of the
algebraic curve. It can be employed in order to set $F$ bigger or
equal to the maximum value of the moduli $b_i$ instead of being smaller 
or equal to the minimum value, thus insuring that $z \rightarrow 0$
corresponds to $F \rightarrow +\infty$ instead of $-\infty$.

The whole procedure is straightforward, but in practice it turns out to
be cumbersome when the moduli parameters $b_i$ take general values.
After all, the uniformization of a genus-9 Riemann surface and the
explicit derivation of the function $u(z)$ is a formidable task.
Matters simplify considerably when one imposes some isometry that
effectively reduces the genus of the algebraic curve as certain 
moduli are allowed to coalesce. In general we will have models
for each continuous subgroup of the maximal isometry group $SO(8)$,
in which case the algebraic curve takes the irreducible form
\be
y^m = (x-b_1)^{a_1}(x-b_2)^{a_2} \cdots (x-b_n)^{a_n} ~, 
\ee
where the integer exponents (with $n \leq 8$) satisfy the relation 
$a_1 + a_2 + \cdots a_n = 2m$. We present below in table 1 all 
Riemann surfaces that classify the domain-wall solutions of 
four-dimensional gauged supergravity with non-trivial scalar fields
in the coset $SL(8,\IR)/SO(8)$ by giving their genus according to
the Riemann--Hurwitz relation, their irreducible form (since in certain
cases the exponents have common factors and the curve might be 
reducible when written in its original form), as well as the corresponding 
isometry groups that determine the geometrical distribution of 
M2-branes in eleven dimensions. We have 22 models in 
total, which are listed.
\begin{center}
\begin{tabular}{c|l|l}
{\hskip-3pt \em Genus} & {\em Irreducible Curve} & {\em Isometry Group}\\
\hline
 9 & $y^4=(x-b_1)(x-b_2) \cdots (x-b_7)(x-b_8)$ & None \\
\hline
 7 & $y^4=(x-b_1)(x-b_2) \cdots (x-b_6)(x-b_7)^2$ & $SO(2)$ \\
\hline
 6 & $y^4=(x-b_1)(x-b_2) \cdots (x-b_5)(x-b_6)^3$ & $SO(3)$ \\
\hline
 5 & $y^4=(x-b_1) \cdots (x-b_4)(x-b_5)^2(x-b_6)^2$ 
& $SO(2) \times SO(2)$ \\
\hline
 4 & $y^4=(x-b_1)(x-b_2)(x-b_3)(x-b_4)^2(x-b_5)^3$ 
& $SO(2) \times SO(3)$ \\
\hline
 3 & $y^4=(x-b_1) \cdots (x-b_4)(x-b_5)^4$ & $SO(4)$ \\
    & $y^4=(x-b_1)(x-b_2)(x-b_3)(x-b_4)^5$ & $SO(5)$ \\ 
    & $y^4=(x-b_1)(x-b_2)(x-b_3)^3(x-b_4)^3$ 
& $SO(3) \times SO(3)$ \\
    & $y^4=(x-b_1)(x-b_2)(x-b_3)^2(x-b_4)^2(x-b_5)^2$ 
& $SO(2) \times SO(2) \times SO(2)$ \\
\hline
 2 & $y^4 = (x-b_1)(x-b_2)^2(x-b_3)^2(x-b_4)^3$ 
& $SO(2) \times SO(2) \times SO(3)$ \\
\hline
 1 & $y^4=(x-b_1)(x-b_2)(x-b_3)^6$ & $SO(6)$ \\
    & $y^4= (x-b_1)(x-b_2)(x-b_3)^2(x-b_4)^4$ 
& $SO(2) \times SO(4)$ \\
    & $y^4=(x-b_1)(x-b_2)^2(x-b_3)^5$ & $SO(2) \times SO(5)$ \\
    & $y^4=(x-b_1)^2(x-b_2)^3(x-b_3)^3$ 
& $SO(2) \times SO(3) \times SO(3)$ \\
    & $y^2=(x-b_1)(x-b_2)(x-b_3)(x-b_4)$ 
& $SO(2)^4$ \\
\hline
 0 & $y^4=(x-b_1)(x-b_2)^7$ & $SO(7)$ \\
    & $y = (x-b)^2$ & $SO(8)$ (maximal) \\
    & $y^2=(x-b_1)(x-b_2)^3$ & $SO(2) \times SO(6)$ \\
    & $y^4=(x-b_1)(x-b_2)^3(x-b_3)^4$ & $SO(3) \times SO(4)$ \\
    & $y^4=(x-b_1)^3(x-b_2)^5$ & $SO(3) \times SO(5)$ \\
    & $y = (x-b_1)(x-b_2)$ & $SO(4) \times SO(4)$ \\
    & $y^2=(x-b_1)(x-b_2)(x-b_3)^2$ 
& $SO(2) \times SO(2) \times SO(4)$ \\
\hline 
\end{tabular}
\end{center}
\begin{center} 
Table 1: Curves and symmetry groups of domain walls 
for M2-branes.
\end{center}
 
We will see later that, for the models with low genus (0 or 1),  
the uniformization can be carried out in all detail and it is possible
in many cases to arrive at explicit expressions for the exact form of
the Schr\"odinger potential.  
     
It is interesting to note that the classification of domain walls 
of five-dimensional gauged supergravity with non-trivial scalar 
fields in the coset $SL(6, \IR)/SO(6)$ follows immediately from 
above by restricting our attention to models with 
an $SO(2)$ isometry factor in 
the symmetry group. It is known that in this case the classification
reduces to the list of all algebraic curves \cite{Baksfe1} 
\be
y^4 = (x-b_1)(x-b_2) \cdots (x-b_6) \ ,
\ee
depending on the values of the six moduli $b_i$. But such curves 
can be viewed as special cases of the $N=8$ curves when
$b_7 = b_8 = \infty$. Therefore, by comparison with  
table 1 we list in table 2 all domain walls of five-dimensional gauged 
supergravity, which correspond to various continuous distributions
of D3-branes in ten dimensions. We have 11 models in
total, which are listed. 
\begin{center}
\begin{tabular}{c|l|l}
{\em Genus} & {\em Irreducible Curve} & {\em Isometry Group}\\
\hline
 7 & $y^4=(x-b_1)(x-b_2) \cdots (x-b_5)(x-b_6)$ & None \\
\hline
 5 & $y^4=(x-b_1)(x-b_2)(x-b_3)(x-b_4)(x-b_5)^2$ & $SO(2)$ \\
\hline
 4 & $y^4=(x-b_1)(x-b_2)(x-b_3)(x-b_4)^3$ & $SO(3)$ \\
\hline
 3 & $y^4=(x-b_1)(x-b_2)(x-b_3)^2(x-b_4)^2$ 
& $SO(2) \times SO(2)$ \\
\hline
 2 & $y^4=(x-b_1)(x-b_2)^2(x-b_3)^3$ 
& $SO(2) \times SO(3)$ \\
\hline
 1 & $y^4=(x-b_1)(x-b_2)(x-b_3)^4$ & $SO(4)$ \\
    & $y^4=(x-b_1)(x-b_2)^5$ & $SO(5)$ \\ 
    & $y^4=(x-b_1)^3(x-b_2)^3$ 
& $SO(3) \times SO(3)$ \\
    & $y^2=(x-b_1)(x-b_2)(x-b_3)$ 
& $SO(2) \times SO(2) \times SO(2)$ \\
\hline
 0 & $y^2 = (x-b_1)(x-b_2)^2$ 
& $SO(2) \times SO(4)$  \\
    & $y^2=(x-b)^3$ & $SO(6)$ (maximal) \\
\hline 
\end{tabular}
\end{center}
\begin{center} Table 2: Curves and symmetry groups of domain walls 
for D3-branes.
\end{center}
 
Finally, the algebraic classification of all domain-wall solutions
of seven-dimensional gauged supergravity with non-trivial scalar
fields in the coset $SL(5, \IR)/SO(5)$ (which by the way provides
the full scalar sector in this case) follows by considering all 
Riemann surfaces of the form
\be
y^4 = (x-b_1)(x-b_2) \cdots (x-b_5)\ ,
\ee
for various values of the five moduli $b_i$. As before, these 
surfaces can be viewed as special cases of the $N=8$ algebraic 
curves where three of the moduli are taken to infinity, i.e. 
$b_6 = b_7 = b_8 = \infty$, whereas the remaining are free to
vary. Put differently, we may compose the list of all domain walls
that correspond to various continuous distributions of M5-branes
in eleven dimensions by considering all $N=8$ models with a 
$SO(3)$ isometry factor. Thus, we have 7 models in total, which 
are listed.
\begin{center}
\begin{tabular}{c|l|l}
{\em Genus} & {\em Irreducible Curve} & {\em Isometry Group}\\
\hline
 6 & $y^4=(x-b_1)(x-b_2)(x-b_3)(x-b_4)(x-b_5)$ & None \\
\hline
 4 & $y^4=(x-b_1)(x-b_2)(x-b_3)(x-b_4)^2$ & $SO(2)$ \\
\hline
 3 & $y^4=(x-b_1)(x-b_2)(x-b_3)^3$ & $SO(3)$ \\
\hline
 2 & $y^4=(x-b_1)(x-b_2)^2(x-b_3)^2$ & $SO(2) \times SO(2)$ \\
\hline
 1 & $y^4=(x-b_1)^2(x-b_2)^3$ & $SO(2) \times SO(3)$ \\
\hline
 0 & $y^4=(x-b_1)(x-b_2)^4$ & $SO(4)$ \\
    & $y^4=(x-b)^5$ & $SO(5)$ (maximal) \\ \hline
\end{tabular}
\end{center}
\begin{center}
Table 3: Curves and symmetry groups of domain walls 
for M5-branes.
\end{center}

In this latter case, the invariance of the curves under the 
discrete symmetry $x \leftrightarrow -x$, $b_i \leftrightarrow - b_i$ 
is not present any more because the algebraic equations contain
an odd number of factors.

\section{Distributions of M2-branes}
\setcounter{equation}{0}

In this section we treat the distributions of M2-branes
with isometries that  correspond to genus 0 and 1 and
present
the uniformization of the associated algebraic curves.
We have 12 such models of low genus, including the $SO(8)$ model
of $AdS_4$ space. When it is possible, we determine
the conformal factor of the metrics (written in conformally flat form), 
as well as the
Schr\"odinger potentials corresponding to the equation 
for the scalar field and graviton fluctuations.
There are several cases where we can explicitly solve the Schr\"odinger
equation and find the 
spectrum by means of elementary methods, 
otherwise we will use the WKB approximation for the computations.

We begin first with the model  
$SO(2)\times SO(2)\times SO(2)\times SO(2)$, which is governed by a
$g=1$ algebraic curve, and study certain limits for which some
of the moduli are let to coincide and the genus is reduced to 0. These
cases are $SO(4) \times SO(2)\times SO(2)$, $SO(4)\times SO(4)$ 
and $SO(6) \times SO(2)$ as well as
the trivial $SO(8)$ model. We also present other $g=0$
models, which cannot be obtained as degenerate limits of the above, 
namely the models with symmetry group $SO(7)$ and $SO(3)\times SO(5)$.
For completeness
we also present the uniformization of the remaining $g=1$ cases,
as well as the remaining $g=0$ model with symmetry $SO(3)\times SO(4)$,
which unfortunately cannot be explicitly brought to a conformally flat 
frame by expressing their uniformizing parameter as a function of $z$,
$u(z)$, in closed form.

\subsection{$SO(2)\times SO(2)\times SO(2)\times SO(2)$}

The algebraic curve can be taken from table 1 and
corresponds to a $g=1$ Riemann surface
\be
y^2 = (x - 1/\lambda_1)(x - 1/\lambda_2)(x-1/\lambda_3) x ~ ,
\ee
with $b_i = 1/\l_i$. We set  $b_4=0$ by a shift of $x$ and
assume the ordering $\l_1 \leq \l_2 \leq \l_3 \leq \l_4 = + \infty$.
Using the birational transformations
\be
y = Y x^2/(2 \sqrt{\lambda_1 \lambda_2 \lambda_3}) ~,\qq
X = -1/x\ ,
\ee
we obtain
\be
Y^2 = 4
(X+\lambda_1)(X+\lambda_3)(X+\lambda_3)~ .
\label{wejfh1}
\ee
Finally, we can bring the curve to its Weierstrass form
\be
w^2 = 4 v^3 - g_2 v - g_3\ ,
\ee
by letting $X = v - (\lambda_1+\lambda_2+\lambda_3)/3$, $Y=w$.  The
coefficients turn out to be 
\ba
g_2 & = & \frac{2}{9} \left((\l_1+\l_2 - 2 \l_3)^2 + (\l_2+\l_3 - 2 \l_1)^2 +
(\l_3+\l_1 - 2 \l_2)^2 \right) ~ , \nonumber \\
g_3 & = & \frac{4}{27}(\l_1 + \l_2 - 2 \l_3) (\l_2 + \l_3 - 2 \l_1) 
(\l_3 + \l_1 - 2 \l_2) ~ . \nonumber
\ea

This elliptic curve can be uniformized in the standard way using 
the Weierstrass functions\footnote{Throught
the rest of the paper we will make use of elliptic functions as
well as of other special functions following the conventions 
of \cite{tipologio,tipolo1,BF}.}
\be
v = \wp(u)\ , \qq w = \wp'(u)  \nonumber ~.
\ee
In general, the two half-periods of the Weierstrass function are given by
\be
\om_1 = {K(k)\ov \sqrt{e_1-e_3}}\ ,\qq \om_2 = {i K(k')\ov \sqrt{e_1-e_3}}
\ ,
\label{wjd1}
\ee
where $K$ is the complete elliptic integral of the first kind with modulus
$k$ and complementary modulus $k'$ given as
\be
k^2 = {e_2-e_3\ov e_1-e_3}\  ,
\qq
k'^2=1-k^2 = {e_1-e_2\ov e_1-e_3}\ . 
\label{wekdfj12}
\ee
Here $e_1$, $e_2$ and $e_3$ are the values of the  Weierstrass function 
at the half-periods, i.e. $\wp(\om_1)=e_1$, $\wp(\om_2)=e_3$ and
$\wp(\om_1+\om_2)=e_2$, which  
are expressed in terms of the parameters $\l_i$ as
\be
e_i = -\l_i +{1\ov 3} (\l_1+\l_2+\l_3)\ ,\qq i=1,2,3\ .
\label{jefh2}
\ee
So, we finally have
\be
F(z g^2) = x = -\frac{1}{\wp(u) - (\l_1+\l_2+\l_3)/3}\ ,
\label{fxpu}
\ee
and
\be
F'(z g^2) = y = \frac{1}{2} (\l_1 \l_2 \l_3)^{-1/2} 
\frac{\wp'(u)}{(\wp(u) - (\l_1+\l_2+\l_3)/3)^2}\ .
\ee

Next we compute $dx/du = 2 (\l_1 \l_2 \l_3)^{1/2} y$ and
$dx/dz = g^2 y$
and find 
\be
u = {g^2 \ov 2} (\l_1 \l_2 \l_3)^{-1/2} z + c\ ,
\ee
where the integration constant $c$ is given implicitly by the relation 
\be
\wp(c) = \frac{1}{3}(\lambda_1 + \lambda_2 + \lambda_3) ~.
\label{pcc}
\ee
The conformal factor is
\be
e^{2 A}  = -\frac{g^2}{2 \sqrt{\l_1 \l_2 \l_3}}\
\frac{\wp'(u)} {(\wp(u) - \wp(c))^2} \ .
\ee
The constant $c$ in \eqn{pcc} was chosen in such a way that $e^{2A}
\sim 1/z^2$ as $z\to 0$
and the space becomes $AdS_4$. Also, according
to the conventions followed in this paper we choose the branch
$\wp'(c)<0$, since $c$ is otherwise determined up to a sign; 
then, the function
$F$ in \eqn{fxpu} indeed approaches $+\infty$ as $z\to 0$.
Using the equation for $c$ and certain identities involving the Weierstrass 
function, the conformal factor can be brought to the form
\be
e^{2 A} = \frac{g^2}{4 \lambda_1 \lambda_2 \lambda_3}
\left( \wp(u-c) - \wp(u+c) \right)\ .
\ee

Finally, the Schr\"odinger potential \eqn{jwd} can be written as
\be
V(z) = \frac{g^4}{4 \lambda_1 \lambda_2 \lambda_3}
\left( 2 \wp(u+c) + 2 \wp(u -  c) - \wp(2 u) \right)\ .
\ee
The structure of the 
spectrum is understood using the results of section 4. For generic
non-identical values of $\l_i$ the spectrum is discrete and the range of 
$0 \leq z \leq  z_{\rm max}$ is finite, where $z_{\rm max}$ is
found by solving the equation $F(z_{\rm max} g^2)=1/\l_1$. We find,
in particular, that
\be
z_{\rm max} = \frac{2 \sqrt{\l_1 \l_2 \l_3}}{g^2} (\omega_1 - c) ~ ,
\ee
where $\om_1$ and $c$ are given by \eqn{wjd1} and \eqn{pcc}.
Then, using \eqn{jkw22} with $D=4$, $\D=4$ and $n=2$, 
we determine the spectrum within the WKB approximation
\be
M_m^2 = \frac{\pi^2}{z_{\rm max}^2} m \left( m+ \frac{1}{2} \right) 
+ {\cal O}(m^0)\ ,\qq m=1,2,\dots \ .
\ee

Note that the model we have just considered corresponds to the 
supersymmetric (extremal) limit of the most general rotating M2-brane 
solution \cite{CVY2} of 
eleven-dimensional supergravity. 
Physically, the parameters $b_i$ are set equal (up to a 
factor) to the
four different rotational parameters that appear in that 
general solution, and therefore are naturally  
associated to the Cartan subgroup of $SO(8)$.
This identification is analogous to the model 
of a D3-brane distribution with $SO(2)\times SO(2)\times SO(2)$ isometry, 
which corresponds
to the supersymmetric limit of the most general rotating D3-brane 
solution \cite{RS1} of type-IIB supergravity,
as shown in \cite{Baksfe1}. 

We will return again to this model in section 8, 
together with other potentials
that arise in elliptic solutions of gauged supergravities, and apply
the rules of supersymmetric quantum mechanics to simplify the 
calculation of the exact spectrum. It will turn out that the partner
potential is formulated as a Lam\'e problem with half-integer 
characteristic
$n=1/2$.
 
\subsection{$SO(4)\times SO(2)\times SO(2)$}

The corresponding algebraic curve of genus 0 is 
$y^2 = (x-b_1)(x-b_2)(x-b_3)^2$, according to table 1. It can be 
brought to the unicursal form $v=w$ by the birational transformation
\be
x={b_1 vw -b_2 \over vw-1} ~, ~~~~ 
y=(b_2 -b_1){v \over (vw-1)^2}\left(b_2 -b_3 -(b_1 -b_3)vw \right) ~.
\ee 
Introducing a uniformizing parameter $u$, so that $u=v=w$, we find after
some computation
\be
dz={2 du \over b_2 -b_3 -(b_1 -b_3)u^2} ~, 
\ee
which can be easily integrated by elementary functions; the final result
depends on the sign of $(b_1 -b_3)(b_2 -b_3)$. 

Alternatively, the result can
be obtained as the limit $\l_1=\l_2$ of the $SO(2)\times SO(2)\times SO(2)
\times SO(2)$ model. 
Using some limiting properties of the Weierstrass functions, 
we find that the conformal factor is  
\be
e^{2 A} = {2 \ov R^2 \l_1 \l_3 } (\l_3/\l_1)^{1/2} (\l_3/\l_1-1)^{3/2}
{\sinh 2 \sqrt{\l_3-\l_1} u \ov (\cosh^2 \sqrt{\l_3-\l_1} u - 
\l_3/\l_1)^2 }\ ,
\label{jwe23}
\ee
and that the Schr\"odinger potential takes the trigonometric form 
\be
V = 4 {\l_3 -\l_1 \ov R^4 \l_1^2\l_3} \left(4 {\l_3/\l_1 
\cosh 2 \sqrt{\l_3-\l_1} u -\cosh^2 \sqrt{\l_3-\l_1} u\ov
 (\cosh^2 \sqrt{\l_3-\l_1} u - \l_3/\l_1)^2 } +1 - {1\ov 
\sinh^2 2 \sqrt{\l_3-\l_1} u}\right)\ , 
\label{jwe11}
\ee
where
\be
u = \frac{g^2 z}{2 \l_1 \sqrt{\l_3}} + c ~ , \qq
c = \frac{\sinh\inv( \sqrt{\l_3/\l_1-1})}{\sqrt{\l_3-\l_1}} \ .
\label{djk21}
\ee
The spectrum is continuous (according to the ordering of our parameters
$\l_1 = \l_2 < \l_3$), and $0\leq z < \infty$. This example 
corresponds to $n=4$, which we have already  
discussed on general grounds in section 4. Hence, there is a 
finite mass gap given by
the value of the potential at $z=\infty$. We find, in particular, that 
\be
M_{\rm gap}^2 = 4 \frac{ \l_3 - \l_1}{\l_1^2 \l_3 R^4} \ .
\ee

Alternatively, one may consider the limit  
$\l_2=\l_3$ of the model with 
$SO(2)\times SO(2)\times SO(2) \times SO(2)$ symmetry.
It turns out that the result is equivalently described by
considering the analytic 
continuation of \eqn{jwe11} and \eqn{jwe13}
by taking $\l_3 < \l_1$. Then, the conformal factor becomes
\be 
e^{2A} ={2 \ov R^2 \l_1 \l_3 } (\l_3/\l_1)^{1/2} (1 - \l_3/\l_1)^{3/2}
{\sin 2 \sqrt{\l_1 - \l_3} u \ov (\cos^2 \sqrt{\l_1-\l_3} u - 
\l_3/\l_1)^2 } \ ,
\label{dfj2}
\ee
and the Schr\"odinger potential is 
\be
V  =  4 {\l_1 -\l_3 \ov R^4 \l_1^2\l_3} \left( 4 {\cos^2 \sqrt{\l_1-\l_3} u
-\l_3/\l_1 \cos 2 \sqrt{\l_1-\l_3} u \ov
 (\cos^2 \sqrt{\l_1-\l_3} u - \l_3/\l_1)^2 } - 1 - {1\ov 
\sin^2 2 \sqrt{\l_1-\l_3} u}\right)\ , 
\label{jwe13}
\ee
where
\be
u  = \frac{g^2 z}{2 \l_1 \sqrt{\l_3}} + c ~ , \qq
c = \frac{\sin\inv ( \sqrt{1-\l_3/\l_1})}{\sqrt{\l_1-\l_3}} \ .
\label{kj211}
\ee
The spectrum is discrete and the range of $z$ is
$0 \leq z \leq z_{\rm max}$, where 
\be
z_{\rm max} = \frac{2 \l_1 \sqrt{\l_3} \sin\inv ( \sqrt{\l_3/\l_1})}
{g^2 \sqrt{\l_1 - \l_3}} \ .
\ee
Using the general results found in \eqn{jkw22} (with $D=4$, $\D=4$, $n=2$), 
we obtain within the WKB 
approximation the spectrum 
\be
\label{ksko}
M_m^2 = \frac{\pi^2}{z_{\rm max}^2} m \left(m+\frac{1}{2}\right) 
+ {\cal O}(m^0)\ ,\qq m=1,2,\dots \ .
\ee


\underline{The exact solution}: 
It might seem hard to find the explicit solution of the Schr\"odinger
equation with potentials \eqn{jwe11} and \eqn{jwe13}.
However, we will show that the explicit solution can be readily  
obtained by applying techniques of supersymmetric quantum mechanics,
thus going beyond the WKB approximation.
We consider for simplicity only 
the case when the spectrum is discrete 
and the potential is given by  \eqn{jwe13}.
It turns out that the partner potential to \eqn{jwe13} is given by 
\be
V_1= {4(\l_1-\l_3)\ov R^4 \l_1^2 \l_3} \left( {3\ov \sin^2 2 \sqrt{\l_1
-\l_3} u}-1\right)\ ,
\label{da11}
\ee
using the notation of section 4, where $u$ is given by \eqn{kj211}.
If we change variables  for the wave-functions of the potential $V_1$ to 
\ba
&& x=\cos^2 \sqrt{\l_1-\l_3} u\ ,\qq 0\leq x \leq {\l_3\ov \l_1}<1\ ,
\nonumber\\
&& \Psi^{(1)} = x^{3/4} (1-x)^{3/4} Y(x)\ ,
\label{he55}
\ea
the new independent variable $Y(x)$ 
satisfies the hypergeometric equation with parameters  
$a= {3/2} + \m$, $b={3/2}-\m$ and $c=2$ in the standard notation, 
where the constant $\m$ is defined
as 
\be 
\m = \ha \sqrt{1 + {R^4 \l_1^2 \l_3\ov 4(\l_1-\l_3) }M^2}\ .
\label{shg11}
\ee 

Hence, we may write down the solution for the wave-function $\Psi^{(1)}$ 
that is regular at $x=0$, as
\be
\Psi^{(1)} \sim x^{3/4} (1-x)^{3/4} F\left({3\ov 2} +\m , {3\ov 2}-\m,2;
x\right)\ .
\label{ggg33}
\ee
Demanding that it vanishes at the end point located at $x=\l_3/\l_1<1$, 
yields the quantization condition for the spectrum 
\be
F\left({3\ov 2} +\m_m , {3\ov 2}-\m_m,2;{\l_3\ov \l_1}\right)=0 \ ,
\qq m=1,2,\dots \ .
\label{qquu}
\ee
It is not possible to find a closed formula for the exact spectrum except for 
the particular case when $\l_1=2 \l_3$. Then, we have $0\leq x \leq \ha$
and making use of the identity
\be
F\left({3\ov 2} +\m_m , {3\ov 2}-\m_m,2;\ha\right)= {\sqrt{\pi}\ov 
\Gamma\left({5\ov 4}+{\m_m\ov 2}\right) \Gamma\left({5\ov 4}-{\m_m\ov 2}
\right)}\ ,
\label{q1u}
\ee
we see that the condition \eqn{qquu} (with $\l_3/\l_1=1/2$) is satisfied
provided that $\m_m=2m +\ha$, where $m=1,2,\dots$. 
Finally, by employing \eqn{shg11}, 
we find that the mass spectrum is given by 
\be
\l_1=2 \l_3 :\qq M_m^2 = {16\ov R^4 \l_3^2} m (m+\ha)\ , \quad m=1,2,\dots\ .
\label{hgh44}
\ee

One may easily check that it coincides with the WKB result \eqn{ksko} (after 
setting $\l_1=2\l_3$), which is also exact in this case. 
The situation is similar to the distribution of D3-branes
with $SO(2)\times SO(4)$ symmetry for which the exact spectrum \cite{FGPW2,BS1}
coincides with what has been found using the WKB approximation \cite{RS1}.
This is a characteristic property of shape-invariant potentials \cite{SQMrev}.
We also note presently that the solution \eqn{ggg33} 
can be expressed in terms of Jacobi polynomials as
\be
\l_1=2 \l_3 :\qq
\Psi^{(1)}_m \sim x^{3/4} (1-x)^{3/4} P^{(1,1)}_{2 m -1}(2 x-1)\ ,
\quad 0\leq x \leq \ha\ ,
\quad m=1,2,\dots \ .
\label{hgh66}
\ee
For the generic case, where the constant parameters $\l_1$ and 
$\l_3$ satisfy the condition $\l_1>\l_3$, but otherwise they remain 
unrelated, 
one may solve \eqn{qquu} in the asymptotic regime of large
$\m$. Using the fact that for large $\m$
\be
F\left({3\ov 2} +\m , {3\ov 2}-\m,2;{\l_3\ov \l_1}\right)\sim 
{\Gamma(\m-1/2)\ov \Gamma(\m+1/2)} \m^{-1/2} (\sin\varphi)^{-3/2}
\sin \m \varphi \left (1 + {\cal O}(\m^{-1})\right)\ ,
\label{jsfh11}
\ee
where the angle $\varphi$ is determined by $\cos\varphi = 1-2\l_3/\l_1$, 
we see that the spectrum is given to leading
order by the solutions of the simple trigonometric equation 
$\sin\m \varphi =0$. 
It follows easily that the result agrees 
with the leading term in the WKB formulae \eqn{ksko}.

Finally, note that 
the wave-functions $\Psi^{(2)}_m$ 
of the corresponding partner potential $V_2=V$ 
have a form that is more complicated than \eqn{ggg33}. They are 
obtained using \eqn{hdh4}, where for the lowering operator 
\eqn{aad2} we have $W(z)= A'(z)$ with $A(z)$ given by
\eqn{dfj2}.
The result is
\ba
\Psi^{(2)}_m &\sim  & 
{\left( x - x^2 \right)^{1/4}\ov
{ \l_1 x - \l_3 }}
\left[
8 \left( \l_3 - 2 \l_3 x + \l_1 x^2 \right)
            \, 
F\left({\frac{3}{2}} + \mu_\m, {\frac{3}{2}} - \mu_m,2,x\right) + \right. 
\nonumber \\
&&\left. + \left( 4 \mu^2_m -9 \right) \left(x - x^2 \right) 
          \left( \l_1 x - \l_3 \right) \,
          F\left(\frac{5}{2} + \mu_m,\frac{5}{2} - \mu_m,3,x\right)
\right]\ ,  
\ea
with $x$ being defined in \eqn{he55}; at the two endpoints
$\Psi^{(2)}_m$ vanishes.

\subsection{$SO(4)\times SO(4)$}

This model is described by an algebraic curve with $g=0$ and can easily
be obtained by taking 
the limit $\l_3\to \infty$ in the above expressions. We find that
the conformal factor for the metric is 
\be
e^{2 A} = {1\ov R^2 \l_1^2} {1\ov \sinh^2\left(2 z\ov R^2 \l_1\right)} \ ,
\ee
whereas the potential is
\be
V = {4\ov R^4\l_1^2}\left(1+{2\ov \sinh^2\left(2 z\ov R^2 
\l_1\right)}\right) \ .
\label{jsf12}
\ee
In this case we have $0 \leq z < \infty$, 
and the spectrum is continuous with a mass
gap
\be
M_{\rm gap}^2 = \frac{4}{R^4 \l_1^2} \ .
\label{ggpp}
\ee

Alternatively, the same result is obtained by direct uniformization
of the irreducible genus 0 curve, $y=(x-b_1)(x-b_2)$ (see table 1).
Letting $v=y/(x-b_1)$, $w=x-b_2$, we arrive at the standard form
$v=w$, which is uniformized by a complex parameter $u$ as
\be
x=u+b_2 ~, ~~~~~ y=u(u+b_2 -b_1) 
\ee
and so the final result reads
\be
z(u)={1 \over b_2 -b_1} \ln\left(u \over u+b_2 -b_1\right) ~; 
~~~~ u(z)={b_1 -b_2 \over 1 - e^{(b_1 -b_2)z}}  
\ee
choosing the integration constant to be zero. Then, we have
\be
y = {(b_1 -b_2)^2 \over 4{\rm sinh}^2 {1 \over 2}(b_1 -b_2)z} ~, 
\ee
which is equivalent to the expression for the conformal factor above,
after introducing the appropriate scale. 

In this model we can also determine the eigenfunctions exactly.
It turns out that the wave-functions are expressed in terms of 
hypergeometric functions, using a characteristic parameter 
$q\neq 0$, as 
\ba
\Psi &=& {1\ov \sqrt{y^2-1}} \left[C_1 y^{1+q} F \left( -\frac{1}{2} 
- \frac{q}{2},
-{q\ov 2},1-q;\frac{1}{y^2} \right) + \right.
\nonumber \\
&& \left. C_2 y^{1-q} F \left( -\frac{1}{2} + \frac{q}{2},
\frac{q}{2},1+q;\frac{1}{y^2} \right) \right]\ , \label{bra22}
\ea
where
\be
y = \cosh \left( \frac{2 z}{R^2 \l_1} \right) ~,\qq  q = \sqrt{1 - 
\frac{M^2 R^4 \l_1^2}{4}} \ ,
\ee
with constant coefficients $C_1$ and $C_2$. A solution 
valid for $q=0$
can also be written down following \cite{tipologio,tipolo1}, but it will not
be needed for the present purposes.
Because of the mass gap \eqn{ggpp}, the 
parameter $q$ is purely imaginary. This provides an  
orthonormalizability in the Dirac sense (with the use of a $\d$-function), 
since the solution \eqn{bra22} becomes an
incoming and an outgoing plane wave for $z\to \infty$.
On the other hand, examining the behaviour of \eqn{bra22} near $z=0$, 
we require the coefficient of the most singular term to vanish.
It yields the following condition, 
\be
C_1 \frac{\Gamma(1-q)}{ \Gamma\left({3\ov 2}-{q\ov 2}\right)\Gamma\left(
1-{q\ov 2}\right)}
+ C_2 \frac{\Gamma(1+q)}{ \Gamma\left({3\ov 2}+{q\ov 2}\right)\Gamma\left(
1+{q\ov 2}\right)}= 0\ .
\ee

Note that the supersymmetric partner of the potential \eqn{jsf12} is
just a constant (given by $V_1={4\ov R^4\l_1^2}$)
and therefore, according to the results of supersymmetric
quantum mechanics, the potential \eqn{jsf12} is reflectionless \cite{SQMrev}.

\subsection{$SO(2) \times SO(6)$}

In this case the irreducible algebraic curve is given by
\be
y^2 = (x - b_1) (x - b_2)^3\ .
\ee
Then, the birational transformation
\be
x  =  \frac{w v b_2 - b_1}{w v - 1} \ ,\qq
y  = - w \frac{(b_1 - b_2)^2}{(w v - 1)^2}\ ,
\ee
brings it to the standard unicursal form $v=w$, 
which is uniformized as usual by introducing a complex parameter $u$,  
$u=v=w$.
We may easily find the coordinate that brings the four-dimensional metric
to a conformally flat form,  
\be
z = - \frac{2(u - 1)}{g^2 (b_1 - b_2)}\ ,
\ee
with conformal factor given by
\be
e^{2 A} = 
{\frac{8\,\left( 2 - {g^2}\,\left( b_1 - b_2 \right) \,z \right) }
   {{g^2}\,{z^2}\,{{\left( 4 - {g^2}\,\left( b_1 - b_2 \right) 
           z \right) }^2}}} \ .
\label{andr1}
\ee

In turn, the Schr\"odinger potential becomes
\be
V(z) =
{\frac{2}{{z^2}}} - {\frac{{g^4}\,{{\left( b_1 - b_2 \right) }^2}}
    {4\,{{\left( 2 - {g^2}\,\left( b_1 - b_2 \right) \,z \right) }^
        2}}} + {\frac{2\,{g^4}\,{{\left( b_1 - b_2 \right) }^2}}
    {{{\left( 4 - {g^2}\,\left( b_1 - b_2 \right) \,z \right) }^2}}
    }
~ .
\label{andr2}
\ee
For $b_1 < b_2$ we have $0\leq z < \infty$ and 
the spectrum is continuous with no gap, which is in 
agreement with our general discussion. Otherwise, for 
$b_2 < b_1$, we find $0 \leq z \leq z_{\rm max} = 2/(g^2 (b_1-b_2))$ and
the spectrum is discrete. 
Using \eqn{jkw22} with $D=4$, $\D=4$ and $n=2$, 
we find within the WKB approximation that the spectrum is 
\be
\label{wkb1}
M_m^2 = \frac{4 \pi^2 (b_1-b_2)^2}{R^4} m \left(m + \frac{1}{2}
\right) + {\cal O}(m^0) ~, \qq m = 1,2,\dots ~.
\ee
Note for completeness that \eqn{andr1} and \eqn{andr2} can also be obtained 
from \eqn{jwe23} and \eqn{jwe11} and from 
\eqn{dfj2} and \eqn{jwe13} by simply taking the limit $\l_3\to \l_1$.

Finally, let us mention that the wave-functions 
and the spectrum can be determined 
exactly using, as before, techniques of supersymmetric quantum mechanics. 
The same is true for the other two cases 
that will be discussed in subsections 6.5 and 6.6 below.
Since there are certain similarities in all three cases of interest, 
we choose to present their analysis based on supersymmetric quantum 
mechanics all together, and uniformly, in subsection 6.7.

\subsection{$SO(3) \times SO(5)$}

The algebraic curve of this model with genus 0 is given by
\be
y^4 = (x - b_1)^3 (x - b_2)^5\ ,
\ee
which by means of the following birational transformation
\be
x  =  \frac{w^3 v b_1 - b_2}{w^3 v - 1} \ ,\qq
y =  -w^4 v \frac{(b_1 - b_2)^2}{(w^3 v - 1)^2}\ ,
\ee
can be brought to the standard form $w=v$, 
which is uniformized (as usual) by a parameter $u$, $u=v = w$.
The coordinate choice that brings the four-dimensional metric
to conformally flat form is
\be
z = {\frac{4(u - 1)}{{g^2}\,\left( b_1 - b_2 \right) u}}\ ,
\ee
and the corresponding conformal factor is 
\be
\label{andr3}
e^{2 A} = 
{\frac{1024\,{{\left( 4 - {g^2}\, b \,z \right) }^3}}
{{g^2}\,{z^2}\,{{\left( 8 - {g^2}\, b \,z \right) }^2}\,
{{\left( 32 - {g^2}\, b \,z\,\left( 8 - {g^2}\, b \,z \right)  
\right) }^2}}}
 ~,
\ee
with $b = b_1 -  b_2$. 

The Schr\"odinger potential turns out to be 
\ba
V(z) &=& 
{\frac{2}{{z^2}}} + \frac{{g^4}\,{b^2}}{4} \,
 \left( {\frac{3}{{{\left( 4 - {g^2}\, b \,z \right) }^2}}} + 
 {\frac{8}{{{\left( 8 - {g^2}\,b\,z \right) }^2}}} - \right. \nonumber \\
 & &  \left. {\frac{512}
 {{{\left( 32 - {g^2}\,b\,z\,\left( 8 - {g^2}\,b\,z \right)  \right) }^2}}} + 
 {\frac{16}{32 - {g^2}\,b\,z\,\left( 8 - {g^2}\,b\,z \right) }} \right) \ . 
\label{jh9}
\ea
For $b_1 < b_2$ the spectrum is continuous without a gap, 
in which case we have
$0 \leq z < \infty$.
For $b_1 > b_2$, we find $0 \leq z \leq z_{\rm max} = 4/(g^2 b)$ and
the spectrum is discrete; within the WKB approximation, it is given by
the simple expression
\be
\label{wkb2}
M_m^2 = \frac{\pi^2 b^2}{R^4} m \left( m+ \frac{3}{2} \right) + 
{\cal O}(m^0)~ , \qq  m = 1, 2, \dots\ ,
\ee
where \eqn{jkw22} has been used with $D=4$, $\D=4$, $n=3$.

\subsection{$SO(7)$}

The corresponding $g=0$ algebraic curve is given by
\be
y^4 = (x - b_1)^7 (x - b_2)
\ee
and by means of the birational transformation
\be
x  =  \frac{w^3 v b_1 - b_2}{w^3 v - 1} \ ,\qq
y  =  -w \frac{(b_1 - b_2)^2}{(w^3 v - 1)^2}\ ,
\ee
it can be brought to the unicursal form $v=w$, 
which is again uniformized as $u=v = w$.
The coordinate that brings the four-dimensional metric
to conformally flat form is
\be
z = -\frac{4(1 - u^3)}{3 g^2 (b_1 - b_2)}
\ee
and the conformal factor is given by
\be
\label{andr4}
e^{2 A} = 
{\frac{g^2 \, b^2 \,
 {{\left( 1 - {3 \ov 4}\,{g^2}\, b \,z 
 \right) }^{{\frac{1}{3}}}}}{\left( -1 + {{\left( 1 - 
 {3 \ov 4} \,{g^2}\, b \,z \right) }^
 {{\frac{4}{3}}}} \right)^2 }} ~.
\ee

The potential of the Schr\"odinger equation of this model is
\be
\label{kjsd11}
V(z) = {\frac{{g^4}\,{b^2}\,
     \left( -5 + 42\,\left( 1 - {3 \ov 4}\,{g^2}\,
           b   \,z \right) \,
        {{\left( 1 - {3 \ov 4} \,{g^2}\, b  \,z
             \right) }^{{\frac{1}{3}}}} + 
       91\,{{\left( 1 - {3 \ov 4} \,{g^2}\, b \,
              z \right) }^{{\frac{8}{3}}}} \right) }{64\,
     {{\left( -1 + {3 \ov 4} \,{g^2}\, b \,z  + 
          {{\left( 1 - {3 \ov 4} \,{g^2}\, b  \,z
                \right) }^{{\frac{7}{3}}}} \right) }^2}}}\ ,
\ee
with $b = b_2 - b_1$. 
For $b_1 > b_2$ we have $0 \leq z  < \infty$ and the spectrum is 
continuous without a mass gap. 
On the other hand, for $b_1 < b_2$, we have $0 \leq z \leq z_{\rm max} = 
4/(3 g^2 b)$ and 
the spectrum is discrete given, within the WKB approximation, 
by 
\be
\label{wkb3}
M_m^2 = \frac{9 \pi^2 b^2}{R^4} m\left(m + \frac{5}{6}\right) 
+ {\cal O}(m^0)~, \qq
 m = 1 ,  2  , \dots\ , 
\ee
where \eqn{jkw22} has been used with $D=4$, $\D=4$, $n=1$.

\subsection{Some exact results}

We are in a position to apply  
techniques of supersymmetric quantum mechanics, as in subsection 6.2
above, in order to find the 
explicit solution of the Schr\"odinger
equation with potentials \eqn{andr2}, \eqn{jh9} and 
\eqn{kjsd11} and determine the associated spectra in a uniform way; 
all three quantum mechanical problems will be treated at once. 

First consider the case where the 
spectrum is discrete. Let us change the variable to $x$, $z=z_{\rm max}-x$, 
where $z_{\rm max}$ is defined as in the appropriate subsections and 
$0\leq x \leq z_{\rm max}$.
Then, by the rules of supersymmetric quantum mechanics,  
the partner potential to our problem is given by 
\be
V_1(x)= {\n^2-1/4\ov x^2}\ ,
\label{hj222}
\ee
where $\n=1,2,\pm 2/3$ for the three models that correspond to 
$SO(2)\times
SO(6)$, $SO(3)\times SO(5)$ and $SO(7)$, respectively.
Both signs $\pm2/3$ were chosen for the $SO(7)$ model for reasons 
that will be indicated towards the end of this subsection.
The solution of the Schr\"odinger equation which is regular at $x=0$
(for $\n=1, 2, 2/3$)
is given in terms of Bessel functions as
\be
\Psi^{(1)} \sim x^{1/2} J_\n(M x)\ .
\label{dj11}
\ee
For $\n=-2/3$ the wave-function diverges at $x=0$, 
but it is not strong enough
to make it non-integrable.\footnote{
In fact, the general criteria developed 
in \cite{wald} render the propagation of quantum test particles 
in such a space-time geometry as unphysical, and hence should be related to 
unphysical vacuum expectation values of the scalar fields. Indeed, it was shown
in \cite{CGLP} that, in this case, the density of the M2-brane distribution has
a negative component which is physically unacceptable.}
Imposing the condition that the wave-function vanishes at $x=z_{\rm max}$,  
which also ensures the 
Hermiticity of the Hamiltonian, we find  
the mass spectrum in terms of the zeros of the Bessel
function, 
\be
J_\n (M_{m} z_{\rm max}) = 0\ ,\qq m=1,2,\dots\ .
\label{zzooo}
\ee
Then, the states in \eqn{dj11}, which  
correspond to the different solutions of \eqn{zzooo}, constitute a complete
set of states. 
The undetermined overall constant in \eqn{dj11} can be found, as usual, 
by demanding the orthonormalizability condition $\int_0^{z_{\rm max}}
dx \Psi^{(1)}_n \Psi^{(1)}_m=\d_{n,m} $ and using the fact that 
\be
\int_0^{z_{\max}} dx x J_\n(M_m x) J_\n(M_n x) ={z^2_{\rm max}\ov 2}
 J_{\n+1}(M_mz_{\rm max} )\d_{m,n}\ .
\label{oorthh}
\ee
Note that the conditions \eqn{zzooo} and $\n>-1$ are crucial 
for the validity of this equation. 

An asymptotic expression for the eigenvalues $M_m$ can be found,
which is valid
for large values of the argument $M_{m} z_{\rm max}$ of $J_\n$.
Using standard formulae from the theory of Bessel functions we  
arrive at the result 
\be
\label{wkb4}
M_m^2 = \frac{\pi^2}{z^2_{\rm max}}m\left(m + \n -\ha\right)
+ {\cal O}(m^0)~, \qq
 m = 1 ,  2  , \dots\ . 
\ee
For $\n=1$ and $\n=2$, it agrees with the WKB formulae \eqn{wkb1} and
\eqn{wkb2}. However, the WKB formula \eqn{wkb3} is reproduced for the value
$\n=-2/3$ (by first shifting $m$ by one unit) instead of $\n=2/3$.
Actually, for $\n=2/3$, 
we obtain \eqn{wkb3} with the number $5/6$ replaced by $1/6$. 
This ambiguity in the spectrum remains unresolved even for  
the partner wave-functions $\Psi^{(2)}$ that we are interested
in computing afterall; 
these are obtained using \eqn{hdh4}, where for the lowering operator 
in \eqn{aad2} we have $W(z)=A'(z)$, with $A(z)$ given by
\eqn{andr1}, \eqn{andr3} and \eqn{andr4} for all three different cases
respectively. 

We find, in particular, for the wave-functions that 
\ba
\n=1 :\quad && \Psi^{(2)}_m \sim  x^{1/2}\left( {2 x\ov 
z^2_{\rm max}-x^2} J_1(M_mx) + M_m  J_0(M_m x)\right)\ ,
\nonumber\\
\n=2 :\quad && \Psi^{(2)}_m\sim  x^{1/2}\left( {4 x^3 \ov 
z^4_{\rm max}-x^4} J_2(M_mx) + M_m J_1(M_m x)\right)\ ,
\label{jjj33}\\
\n=\pm {2\ov 3} :\quad &&\Psi^{(2)}_m \sim  { x^{1/2}\ov
z_{\rm max}^{4/3}-x^{4/3}} \left( 
x^{4/3} J_{\pm 5/3}(M_mx) + z_{\rm max}^{4/3} J_{\mp1/3}(M_m x)\right)\ .
\nonumber
\ea
It is crucial to note here that for all three type of wave-functions 
above, $\Psi^{(1)}_m$
vanish at the end point $x=z_{\rm max}$.
This can be shown by first 
expanding the wave-function around $x=z_{\rm max}$, 
using properties of the Bessel functions, and then observe that the 
coefficient of the divergent part in the expansion is proportional 
to $J_\n(M_m z_{\rm max})$, and hence vanishes due to \eqn{zzooo};
the constant part vanishes identically.
Note also that the asymptotic behaviour of the wave-function 
near $x=0$ is $\Psi^{(1)}_m\sim x^{1/6}$ for 
both $\n=2/3$ and $\n=-2/3$.
Hence, there is a priori no reason to dismiss either 
one of the two values $\n=2/3$
or $\n=-2/3$. 
It remains unclear to us, at least for the moment being, 
what is the extra condition 
one should impose in order to exclude one of these two values.  
As soon as this becomes possible, 
the spectrum corresponding to 
the distribution of M2-branes with $SO(7)$ symmetry will be determined 
unambiguously in the discrete case.

Considering the same potentials 
\eqn{andr2}, \eqn{jh9} and \eqn{kjsd11}, but for continuous  
spectrum, we find in all three cases that   
the parameter $z_{\rm max}$ becomes negative.
By changing variable to $x$, where $z= x-|z_{\rm max}|$, 
we note that the appropriate partner potential is still given by \eqn{hj222},
but with $|z_{\rm max}|\leq x < \infty$. 
Since the point $x=0$ is not
contained in the range of $x$ we should admit both independent solutions of the
corresponding Bessel equation, namely $J_\n$ and $N_\n$, following the
standard nonmenclature. Demanding that 
the wave-functions vanish at the endpoint $x=|z_{\rm max}|$, 
determines their
relative coefficient. Hence, the wave-function is given by 
\be
\Psi^{(1)} \sim N_\n(M |z_{\rm max}|) J_\n(M x) - 
J_\n(M |z_{\rm max}|) N_\n(M x) \ ,
\label{dj9}
\ee
whereas the spectrum is continuous with no mass gap.

\subsection{$SO(3) \times SO(4)$ }

Last, but not least, we consider the remaining curve of genus 0 
\be
y^4 = (x - b_1) (x - b_2)^3 (x-b_3)^4 ~,
\ee
which can be brought to the unicursal form using the birational
transformation
\be
x= {b_1 - b_2 v^3 w \over 1 - v^3 w} ~, ~~~ 
y = {(b_1 - b_2) v \over 1 - v^3 w} \left( {b_1 - b_2 v^3 w 
\over 1 - v^3 w} - b_3 \right) 
\ee
and hence uniformized by setting $u = v = w$. Assuming for
definiteness  that $b_1 \neq b_2$, 
we may proceed to solve
the differential equation \eqn{ssool} in order 
to obtain the corresponding function
$z(u)$. Taking the limits $b_1 \to b_3$ or $b_2 \to b_3$
yields the models $SO(3) \times SO(5)$ and
$SO(7)$ respectively, which we have already discussed. 
For general $b_i$'s we arrive at
\ba
\alpha <0  : \qq
z & = & \frac{1}{\sqrt{2} g^2 (b_1-b_3) q^{3}}
\left[
2 \tan\inv \left(\frac{\sqrt{2} q u}{1 - q^2  u^2}\right)
  + \right. \nonumber \\
&& \left.
+ \ln\left( \frac{1 - \sqrt{2} q u 
+ q^2 u^2}{1 + \sqrt{2} q u + q^2 u^2} \right) \right]
+ {\rm const.}\ , \\
\alpha >0  : \qq
z & = & \frac{1}{g^2 (b_1-b_3)q^{3}}
\left[ -2 \tan\inv q u + \ln \left(
\frac{1+q u}{1-q u}\right)  \right] +  {\rm const.}\ ,
\nonumber
\ea
with
\be
\alpha = \frac{b_2-b_3}{b_1-b_3} \  , \qq  q = |\alpha|^{1 \ov 4}\ .
\ee

Unfortunately, it is not possible to invert the relations and find
$u(z)$ in closed form, and so they will not be pursued any further.

\subsection{More $g=1$ surfaces}

In the following we present the uniformization of the
remaining models with genus 1 in a unifying way.
These cases are
\ba
(i) & y^4 = (x-a) (x-b) (x-c)^6 \ \ \ & :  \ \ SO(6)\ , \nonumber \\
(ii) & y^4 = (x-a) (x-b) (x-c)^2 (x-d)^4 \ \ \ & :  \ \ SO(2) \times SO(4) \ ,
\nonumber \\
(iii) & y^4 = (x-a) (x-c)^2 (x-b)^5 \ \ \ & :  \ \ SO(2) \times SO(5) \ ,
\nonumber \\
(iv) & y^4 = (x-c)^2 (x-a)^3 (x-b)^3 \ \ \ & :  \ \ SO(2) \times SO(3) 
\times SO(3) \ .
\nonumber 
\ea
Using birational transformations, they can
be brought into the same form
\be
\label{common}
(X-c)^2 Y^4 = (X-a)(X-b)\ ,
\ee
where in each case we consider the following:  
\ba
(i)   & & X = x \  , \ \ Y = {y \ov (x-c)^2} \ ,   \nonumber \\
(ii)  & & X = x \  , \ \ Y = {y \ov (x-c)(x-d)} \ , \nonumber \\
(iii) & & X = x \  , \ \ Y = {y \ov (x-b)(x-c)} \ , \nonumber \\
(iv)  & & X = x \  , \ \ Y = {(x-a)(x-b) \ov y}\ \nonumber .
\ea
Then, the birational transformations are employed 
\ba
X & = & c \frac{v^2 + \frac{k}{2} \frac{2 a b - c(a+b)}{c(a-b)} v +
\frac{k^2}{16} }
{v^2 + \frac{k}{2} \frac{a + b - 2 c}{a - b} v +
\frac{k^2}{16}} \ ,
\nonumber \\
Y & = & \sqrt{\frac{b-c}{a-c}} \frac{w}{2 v} \ , \ \ {\rm with} 
\ \ k^2 = \frac{(a-b)^2 (a-c)}{(b - c)^3}\ ,
\label{biratio}
\ea
to bring the common form \eqn{common} into the Weierstrass
normal form of the curve 
\be
w^2 = 4 v^3 - g_2 v - g_3 \ ; \qq g_2 = \frac{k^2}{4} \ ,\qq g_3 = 0 \ .
\ee

Note that in all cases we have set 
$x = X$, with $X$ given by \eqn{biratio} above, whereas for $y$ we have to
treat each transformation separately. 
Explicit calculation shows that $y$ is 
equal to 
\ba
(i) & & \frac{(a-c)^3}{\sqrt{(a-c)(b-c)}} \frac{v w}{2} 
\frac{1}{\left( v^2 + \frac{k}{2} \frac{a + b - 2 c}{a - b} v +
\frac{k^2}{16} \right)^2}\ ,
\nonumber \\
(ii) & & \frac{(a-c)^2}{\sqrt{(a-c)(b-c)}} \frac{w}{2} 
\frac{(a-d) v + (c-d) \sqrt{\frac{b-c}{a-c}} (v + k/4)^2  }
{\left( v^2 + \frac{k}{2} \frac{a + b - 2 c}{a - b} v +
\frac{k^2}{16} \right)^2}\ ,
\nonumber \\
(iii) & & {\rm same \ as \ in} \ (ii) \ {\rm setting} \ d = b \ ,
\nonumber \\
(iv) &  & -\frac{(a-c)^2}{(b-c)} \frac{2 v (v + k/4)^2}{w} 
\frac{(a-b) v + (c-b) \sqrt{\frac{b-c}{a-c}} (v + k/4)^2  }
{\left( v^2 + \frac{k}{2} \frac{a + b - 2 c}{a - b} v +
\frac{k^2}{16} \right)^2} \nonumber
\ea
As usual, in all four models we have to solve the differential equation
\eqn{ssool} in order to determine the function $u(z)$. 

Using $v = \wp(u)$ and $w = \wp'(u)$ we find the following results in 
each case separately:
\ba
(i)   & & z = {4 \over a-c} \left( \zeta(u) + {1 \over 4} 
{\wp'(u) \over \wp(u)} \right) + {\rm const.} \nonumber \\
(ii)  & & z = - \frac{8 (b-c)^3}{(c-d)(a-b)^2} \frac{1}{\sqrt{(a-c)(b-c)}} 
\left[ {1 \over 4} k^2 u +  \right. \nonumber   \\
 & & \left. + \frac{v_- \wp'(a_+)}{v_+ - v_-} \left( \log 
{\sigma (u-a_+) \over \sigma(u+a_+)} + 2 \zeta(a_+) u \right) - \right. \\ 
 & & \left. - \frac{v_+ \wp'(a_-)}{v_+ - v_-} \left(\log 
{\sigma (u-a_-) \over \sigma (u+a_-)} + 2 \zeta(a_-) u  
\right)  \right] + {\rm const.} \ , \nonumber \\ 
(iii) & & z = -\frac{4}{a-b} \left( 2 \zeta(u) +  
{\wp'(u) \over \wp(u) - k/4} \right) +
{\rm const.} \nonumber \\
(iv) & & z = -{2 \over b-c} u  
+ {\rm const.} ~, \nonumber
\ea
where $v_{\pm}$ are the two roots of the equation 
\be
(v+k/4)^2 +
\frac{a-d}{c-d} \sqrt{\frac{a-c}{b-c}} v =0 
\ee
and $a_{\pm}$ are defined by $v_{\pm} = \wp(a_{\pm})$. In the case (ii) above
we assume that $c \neq d$, and generically $a_{\pm}$ differ from the 
half-periods of the corresponding Riemann surface. Although  
case (i) corresponds to taking $c=d$, there is no smooth limit of (ii) 
that yields the expression (i). 
In the case (iii) above  
we have $v_+ = v_-$, and so the derivation has to be performed separately
without taking (ii) in the limit $b=d$. 

It is rather unfortunate that we can not  
invert the relations and find $u(z)$ in closed form for these models, 
apart from the case (iv). Hence, they
will not be discussed any more. We only leave case (iv) as an 
exercise for the interested reader to explore it further.

\section{Distributions of M5-branes}
\setcounter{equation}{0}

In this section we treat the distributions of M5-branes 
with isometries that  correspond to genus 1 and 0 and
present
the uniformization of the associated algebraic curves in 
as much the same way
as for the M2-branes. 
There are only two models to consider apart from the $AdS_7$ space, 
namely the distribution with isometry group  
$SO(2)\times SO(3)$ that corresponds to a genus 1 algebraic curve,
and that with isometry group $SO(4)$ corresponding to a genus 0 algebraic 
curve.
Unfortunately there is no case where 
we could find the exact spectrum in terms of known functions and therefore 
we will only resort to the WKB approximation for the computations.

\subsection{$SO(2) \times SO(3)$}

Consider the genus 1 curve of table 3 
\be
y^4 = (x - a)^2 (x - b)^3 ~.
\ee
With the aid of the birational transformation 
\be
x  =  \frac{b - a}{4} \frac{w^2}{v^3} + a \ , \qq
y  =  (b-a) \frac{w}{v} \left(\frac{w^2}{4 v^3} -1 \right) \ ,
\ee
it can be brought into the Weierstrass form
\be
w^2 = 4 v^3 - \frac{a-b}{4} v\ ,
\ee
which is uniformized using the Weierstrass functions defined for 
$g_2 = \frac{a-b}{4}$ and $g_3 = 0$.
Since  
\be
x  =  b + \frac{(a - b)^2}{16 \wp(u)^2} \equiv F(z g^2) \ ,\qq
y  =  \frac{(a - b)^2 \wp'(u)}{16 \wp(u)^3} \equiv F'(z g^2)\ ,
\ee
we obtain  
\be
u = c - \frac{g^2 z}{2} \ , 
\ee
where $c$ is an integration constant (to be fixed by the asymptotic 
conditions). Then, using \eqn{ai10}, we find the conformal factor 
\be
e^{2 A} = g^2 F'(z g^2)^\frac{2}{5} ~.
\ee

Note that the two half-periods of the torus $\omega_1$, $\omega_2$ 
can be computed using \eqn{wjd1} and the fact that in our model the modulus
and its complement are equal to each other, $k=k'=1/\sqrt{2}$, 
since $g_3 = 0$. We have 
\be
\om_1 = -i \om_2 ={ \G(1/4)^2\ov 2 \sqrt{2 \pi} (a-b)^{1/4}}\ .
\label{jd44}
\ee
Hence, for $a>b$, $\om_1$ is real and $\om_2$ is purely imaginary, 
whereas for $a<b$ they are complex conjugate of each other. 
The constant of integration $c$ is determined 
by requiring that the space becomes $AdS_7$ for $z\to 0$. 
We find, in particular, that $c=\om_1+\om_2$, which is complex
(or real) if $a>b$ (or $a<b$).

We also find the following Schr\"odinger potential, using the 
rescaling factor $g=2/R$,  
\be
V(z)  =  \frac{1}{R^4} \left[ 35 \wp \left(\frac{2 z}{R^2} \right) +
3 \wp \left(\frac{2 z}{R^2} + \omega_1 + \omega_2  \right) 
 -\wp \left(\frac{2 z}{R^2} + \omega_1 \right) -
\wp \left(\frac{2 z}{R^2} + \omega_2 \right) \right] \ .
\nonumber
\label{jfh5}
\ee
Note that $0\leq z < z_{\rm max}$, where 
$z_{\rm max}=\frac{R^2}{2} \left( \omega_1 + \omega_2 \right)$ if $a<b$ 
and $z_{\rm max}=\frac{R^2}{2} \omega_1 $ if $a>b$.
The spectrum is discrete and, within the WKB approximation, it is given by 
\ba
&& a< b : \qq M_m^2 = \frac{16 \pi^3 \sqrt{b-a}}{\G(1/4)^4 R^4 } m (m+3)
+ {\cal O}(m^0) \ ,\qq  m=1,2,\dots \ ,
\label{wj223}
\\
&& a> b : \qq M_m^2 = \frac{32 \pi^3 \sqrt{a-b}}{\G(1/4)^4 R^4} m (m+2)
+ {\cal O}(m^0) \ ,\qq  m=1,2,\dots \ , 
\label{wj213}
\ea
where \eqn{jkw22} has been used with $D=7$, $\D=4$ and $n=3$ (or $n=2$)
for $a<b$ (or $a>b$).
We will see later, in the context of supersymmetric quantum mechanics,
that the partner potential is related to the potential of 
the $SO(3) \times SO(3)$ model of D3-branes in five dimensions. 

\subsection{$SO(4)$}

Consider next the genus 0 curve 
\be
y^4 = (x-a) (x-b)^4\ .
\label{hho}
\ee
The birational transformation
\be
x = \frac{1}{v^3 w} + a \ , \qq y = \frac{1}{v} \left( \frac{1}{v^3 w} +
a - b \right) \ ,
\ee
brings it into the unicursal form $v=w$, 
which can be uniformized with a complex parameter $u$, as $ v = w=u$.
Consequently, we arrive at
\be
dz = -{4 \ov g^2} \frac{du}{(a-b) u^4 + 1}\ ,
\ee 
which yields upon integration the following cases:
\ba
 a=b : && z = -{4 u \ov g^2} + {\rm const.} \ , \ 
{\rm which \ gives} \ AdS_7\ , 
\nonumber \\
 a<b  :&&  z = -\frac{1}{g^2 \sqrt[4]{b-a}} \left[ \ln\left(
\frac{1 + u \sqrt[4]{b-a}}{1 - u \sqrt[4]{b-a}}\right) + 
2 \tan\inv (u \sqrt[4]{b-a}) \right] + {\rm const.}\ ,  \\
 a>b : && z = -\frac{1}{g^2 \sqrt{2} \sqrt[4]{a-b}} \left[
\ln \left( \frac{1 + u \sqrt{2} \sqrt[4]{a-b} + u^2 \sqrt[4]{a-b} }
{1 - u \sqrt{2} \sqrt[4]{a-b} + u^2 \sqrt[4]{a-b}} \right)
+ \right. \nonumber \\
 && \qq + \left.  
2 \tan\inv \left(\frac{u \sqrt{2} \sqrt[4]{a-b}}{1 - u^2 \sqrt{a-b}} \right)
\right] + {\rm const.}\ . \nonumber
\ea
Apart from the maximally symmetric model that corresponds to 
$a=b$, these relations cannot be inverted to yield $u(z)$ in 
closed form.

The nature of the spectrum depends crucially on the sign of $a-b$.
Using our general formulae we find that 
for $ a< b$ we have $0\leq z < \infty$ and that the spectrum is 
continuous with a mass gap given by 
\be 
a<b : \qq M_{\rm gap}^2 = 4 {\sqrt{b-a}\ov R^4}\ .
\label{w5fh1}
\ee
For $a> b$ we have $0\leq z \leq z_{\rm max}=\sqrt{2}\pi / g^2
(a-b)^{1/4}$ and the spectrum is discrete. It is approximated by
the WKB formulae \eqn{jkw22} with $D=7$, $\D=4$ and $n=1$ as 
\be
a>b : \qq M_m^2 = \frac{8 \sqrt{a-b}}{R^4} m (m+\frac{7}{3}) 
+ {\cal O}(m^0)\ ,\qq m=1,2,\dots \ .
\label{dj22}
\ee

Finally we would like to mention the relation of the models we have presented 
in this section to the most general solution of 
rotating M5-brane \cite{CRST}
of eleven-dimensional supergravity.
The latter, besides the usual Poincar\'e invariance along the brane,
has also an $SO(2)\times SO(2)$ symmetry group corresponding to the Cartan 
subgroup of $SO(5)$. Hence, in the extremal limit it will correspond
to a supersymmetric solution associated with an algebraic curve of genus 2, 
as can be seen from the appropriate entry in table 3.
The two independent parameters in the equation of the algebraic curve are 
related to the rotational parameters of this rotating M5-brane solution.
The genus 1 model 
with symmetry $SO(2) \times SO(3)$ corresponds to
the particular limit when one rotational parameter is set equal to zero.
The case with $a<b$ corresponds to the rotating solution with 
Lorentzian signature, whereas for $a>b$ 
it corresponds to the same solution, but with the
time and angular parameters analytically continued so that the metric 
remains real but its signature becomes Euclidean.
The associated spectra are given in the WKB approximation 
by \eqn{wj223} and \eqn{wj213}.\footnote{In fact \eqn{wj213} coincides 
with the supersymmetric limit of
the WKB formula given in \cite{minahan} for the masses of
$0^{++}$ glueballs using rotating M5-branes with one rotational parameter,
in a supergravity approach to QCD$_4$ \cite{Witten}.}
The genus 0 model with symmetry group $SO(4)$ 
corresponds to letting the two angular parameters become
equal. We note that this 
is not equivalent to setting one of them zero and keeping the other  
finite. Then,  $a<b$ describes the Euclidean solution, whereas  
$a>b$ describes the Lorentzian. 
The corresponding spectra are described by
\eqn{w5fh1} and \eqn{dj22} respectively.

\subsection{Wilson surfaces}

We would like to calculate the vacuum expectation values of
Wilson surface operators in the six-dimensional $(0,2)$ theories on
the Coulomb branch. It was shown in \cite{wilsonsurf} that the AdS/CFT
correspondence could be used to compute Wilson surface observables 
\cite{ganor} of $(0,2)$ theories in the limit of a large number $N_b$ of
M5-branes. 
The Wilson area operator in the supergravity picture is defined by
requiring that a membrane ends at the boundary of $AdS_7 \times S^4$
on the surface that defines the operator. We will consider Wilson
surfaces 
corresponding to a pair of parallel strings on the boundary
using the prescription of \cite{wilsonsurf} in the special backgrounds 
constructed in sections 3.1, 7.1 and 7.2.
Wilson loops turned out to be useful
tools for learning about the physics of gauge theories
in the study of supergravity duals of four-dimensional theories
on the Coulomb branch. 
It is interesting that complete screening was found with an associated
screening length suppressed by $1/\sqrt{g_s N_b}$ compared
to what is expected from field theory considerations at weak coupling
\cite{FGPW2,BS1}. 

In the conformal limit, 
this calculation was performed in \cite{wilsonsurf}
leading to the result
\be
\label{surfpot}
E = -\frac{8 \sqrt{\pi} \Gamma \left( {2 \ov 3} 
\right)^3}{\Gamma \left( {1 \ov 6} \right)^3} \frac{N_b}{L^2}\ ,
\ee
where 
$E$ denotes the energy per unit length or tension between the infinitely 
long strings as function of their separation $L$.  
This can easily be generalized to backgrounds of the form \eqn{M55}, 
\eqn{ejh1}-\eqn{dhj1} 
with $f$ defined as in equation \eqn{hj3}. 
We will impose the minor restriction 
$b_1 = \ldots = b_n = b$ and $b_{n+1} = \ldots = b_5 = 0$, thus 
breaking the $SO(5)$ symmetry to $SO(n) \times SO(5-n)$. Furthermore,
we choose the orientation of the Wilson surface on the deformed $S^4$ 
to be constant and to lie in the subspace spanned by 
$y_{n+1}, \ldots, y_5$ (see also \eqn{ejh1}). 

Minimizing the membrane action in these backgrounds 
with the orientation chosen as above, yields the 
following integrals for the length and the energy (for $1 \leq n \leq 4$)
\ba
L & = & \frac{R^2}{2} \int_{F_0}^\infty dF \frac{\sqrt{h(F_0)
g(F)}}{\sqrt{
h(F) (h(F)-h(F_0))}} \ \ ,\\
E & = & \frac{1}{\pi^2 R} \int_{F_0}^\infty dF \frac{\sqrt{h(F) g(F)}}
{\sqrt{h(F)-h(F_0) }} -
\frac{1}{\pi^2 R} \int_{b}^\infty dF \sqrt{g(F)} \ \ , \\
h & = & (F-b)^\frac{n}{2} F^\frac{3-n}{2} \ \ , \ g \ = \ F^{-1} \ \ ,
\ea
with $F_0 \geq b$ being the minimal value of $F$ that  
the Wilson surface reaches. In general 
these integrals cannot be expressed in terms of 
known functions and so 
we will only present here some numerical results.
For $F\gg b$, or equivalently for small separations 
$L\ll R^2/\sqrt[4]{b}$, the behaviour of the potential is as in
\eqn{surfpot}
and goes to zero faster for larger separations. From a certain distance
$L > L_{\rm max}$, and further on, 
there does not exist a minimal surface connecting
the two strings on the boundary. Instead, a configuration of two separated
surfaces hanging straight into the interior of the geometry is 
energetically preferred. This means that the 
potential is screened 
for large separation; a phenomenon that was also observed in
four-dimensional
superconformal theories on the Coulomb branch \cite{FGPW2,BS1}. 
The maximal
distance at which the string breaks can be determined numerically
\be
L_{\rm max} = c_n \frac{R^2}{ \sqrt[4]{b} } 
\ee
with $c_1 \sim 0.71$ , $c_2 \sim 0.69$ , $c_3 \sim 0.70$ and $c_4
\sim 0.78$. For $n = 1, ~2, ~3$ the length reaches its
maximum $L_{\rm max}$ at $F_0 > b$ and becomes zero as $F_0 \to b$.
At $L = L_{\rm max}$ the energy is larger than zero and the split
configuration is preferred. For $n=4$, $L_{\rm max}$ is reached
exactly when $F_0 \to b$; at this point the potential tends 
smoothly to zero and will remain there even if the separation is
increased. This is the phenomenon of complete screening that was also
found 
in the context of some special continuous distributions of D3-branes in
\cite{FGPW2,BS1}. This phenomenon occurs in cases where the mass spectrum
is continuous with a mass gap.

\section{Comments on Lam\'e equations}
\setcounter{equation}{0}

In this section we summarize some results on the Lam\'e equation, 
and its various generalizations, which arise in the study of quantum
fluctuations for the scalar and graviton fields in the background of 
domain walls associated to elliptic functions. We have already seen
that in many cases the Schr\"odinger potential has the common form  
\be
V(u) = \l(\l+1) \wp(u)+ \m(\m+1)\wp(u+\om_1)+ \n(\n+1)\wp(u+\om_1+\om_2)
+\kappa(\kappa+1) \wp(u+\om_2)\ ,
\label{uunivv}
\ee
for various choices of the coefficients $\l$, $\m$, $\n$ and $\kappa$; they
are all constrained to satisfy the Hermiticity bound $\geq -1/2$. We have
found, in particular, the following list of examples:  
\ba
 (i)\  &&
{\rm D3\!-\!branes\ with}\ SO(2)\times SO(2)\times SO(2): \qq \l={3\ov 2}\ ,
\quad \m=\n=\kappa =-\ha\ ,
\nonumber\\
 (ii)\  && {\rm D3\!-\!branes\ with}\ SO(3)\times SO(3): 
\qq \l=\n={3\ov 2}\ ,
\quad \m=\kappa =\ha\ ,
\nonumber\\
 (iii)\ && {\rm M5\!-\!branes\ with}\ SO(2)\times SO(3): \qq \l={5\ov 2}\ ,
\quad \n=\ha\ ,\qq \m=\kappa =-\ha\ .
\nonumber
\ea
The first two cases were derived in \cite{Baksfe1}, where emphasis
was placed on analyzing 
domain wall solutions of five-dimensional gauged 
supergravity.\footnote{Actually, for the $SO(3) \times SO(3)$ 
model of D3-branes, the Schr\"odinger potential was originally presented 
in another form in \cite{Baksfe1}, but that is equivalent to the 
potential (ii) because of special identities of 
the underlying
Riemann surface with $g_3 = 0$.}
There, the expressions for the conformal factor were  
found to be 
\be
e^{2A(z)} = \left( \wp^{\prime}(u)\ov 2 R^3\right)^{2/3} ~~~~
{\rm and} ~~~ \left({\wp^{\prime}(u) \over 4 R \wp(u)}\right)^2 ~,
\ee
with $u = z/R^2$ and $u = z/(2R^2)$, respectively.
The third example was discussed here in section 7.1, and has 
$u = 2z/R^2$.
The parameter $u$ assumes real values
from $0$ to $\omega_1$ (real semi-period) 
for the cases (i), (ii) and (iii) (with 
$a>b$), whereas for the case (iii) (with $a<b$) $u$ takes real values from 
$0$ to $\om_1+\om_2$.
In the context of supersymmetric quantum mechanics \cite{SQMrev}  
the potential (i) is mapped to a potential with
coefficients $\l=1/2$, $\m=\n=\kappa=0$, which is simpler to study. 
The potentials (ii) and (iii) (with $a>b$) also turn out to be related to each
other via supersymmetric quantum mechanics, and the details are 
worth exposing. 

More precisely, 
we find that the supersymmetric
partner potentials corresponding to the $SO(2) \times SO(2) \times SO(2)$ 
model in $D=5$ are given respectively by the pair 
\be
V_1(u) = 3 \wp(2u) ~,\qq V_2(u) = 4 \wp(u) - \wp(2u) \ ,
\ee
making use of the identity
\ba
\wp(2u) & = & -2 \wp(u) + \left({\wp^{\prime \prime}(u) \over 
2 \wp^{\prime}(u)}\right)^2 \nonumber\\
& = & {1 \over 4} \left( \wp(u) + \wp(u + \omega_1) + 
\wp(u + \omega_2) + \wp(u + \omega_1 + \omega_2) \right) ~. 
\ea
So, by rescaling $u$ by a factor of 2, setting $\tilde{u} = 2u$, we obtain 
a partner Schr\"odinger problem in $\tilde{u}$ with potential 
$V_1(\tilde{u}) = \lambda (\lambda +1) \wp(\tilde{u})$ having  
$\lambda = 1/2$ as advertised. For later use we drop the tilde and
still use the variable $u$, but with a range from 0 to $2 \omega_1$.
According to the previous general discussion, the ground state energy of 
$H_1$ is expected to be zero, 
which will also be encountered later using a direct
approach.

The Schr\"odinger potentials appearing in the examples (ii) and (iii) (with
$a>b$)  
also have supersymmetric partners within the same class. We find, 
in particular, that the supersymmetric partner potential of the 
model (ii) has
$\lambda = \nu = 1/2$, $\mu = \kappa = 3/2$, whereas a similar
analysis for the model (iii) yields a potential with 
$\lambda = \nu = 3/2$, $\mu = \kappa = 1/2$. 
We do not observe important simplifications occuring  
in the form of the potential in any of the two cases. An 
interesting observation is that the 
supersymmetric partner of the potential (iii) is (ii), and we
believe that there is a deeper reason for this finding.
Note at this point that, generically, any two supersymmetric 
partner potentials are related to each other by simply changing
sign in the superpotential $W \leftrightarrow -W$. For domain wall
solutions this would mean that the conformal factor reverses, since 
$A \leftrightarrow -A$, and so this transformation could only be of
mathematical interest in relating different spectra. In physical
terms, the transformation $W \leftrightarrow -W$ cannot be
used to map one
domain wall solution to another because it fails to preserve the 
AdS boundary condition imposed on the conformal factor 
for $z \rightarrow 0$. However, it may happen in certain cases (as
above) that there are two different superpotentials with the 
correct asymptotic behaviour as $z \rightarrow 0$ which yield
the same supersymmetric partners; indeed, the potential (ii) is the $V_1$ 
partner of a potential $V_2$ given by the model (iii), due to special
identities on Riemann surfaces with $g_3 =0$, but conversely this is
not so because $W \leftrightarrow -W$ does not relate the model
(ii) and (iii). 

Another notable relation concerns the supersymmetric partner of the
elliptic potential for the $SO(2) \times SO(2) \times SO(2) 
\times SO(2)$ model of M2-branes in four-dimensional gauged 
supergravity. Actually, in this case we find that the two
partner potentials
\be
V_1 = 3 \wp(2u) ~, ~~~~ 
V_2 = 2\wp(u+c) +2\wp(u-c) -\wp(2u) 
\ee
are connected by supersymmetry, and so by reinstating the overall
scaling factor and constant shift that relates the uniformizing parameter
to the Schr\"odinger variable (now called $u$), we obtain again 
the Lam\'e potential with $\l = 1/2$ but in the range from 
$2c$ to $2 \omega_1$.   
This concludes the 
presentation of some qualitative results on the Schr\"odinger 
equation of 
quantum fluctuations on elliptic backgrounds.    

An interesting problem that remains unsolved is the exact evaluation of the 
full spectrum of the Schr\"odinger equation in this class of potentials.
So far we have relied on semi-classical approximation methods to get a
feeling about the spectrum and the existence of a mass gap.  
Unfortunately, the exact result is very hard to find, even in some
simple cases, and involves transcendental equations in rather implicit
form. Nevertheless, it is quite instructive to highlight some special
results in order to appreciate the degree of difficulty one  
faces in the general case. We note that the family of potentials under
investigation are indeed a natural generalization of Lam\'e's potential
$n(n+1) \wp(u)$ by adding terms located at all four corners of the 
parallelogram $u$, $u + \omega_1$, $u + \omega_2$, 
$u + \omega_1 + \omega_2$ in the complex domain of a genus 1 Riemann
surface. The Lam\'e potential was originally introduced in order to 
describe the analogue of spherical harmonics for the solutions of 
the Laplace equation in three dimensions with ellipsoidal symmetry. As
such, $n$ can only take (positive) integer values. However, 
generalizations were also considered, more than a century ago, for 
half-integer and other values of the parameter $n$. The main results
in this direction go back to Hermite, Brioschi and Halphen (see, for
instance, \cite{Ref3} and \cite{ince}), 
but also Darboux apparently
studied some aspects of the general potential \eqn{uunivv} in its
Jacobi form. Half-integer values of the coefficients are particularly
interesting for the examples we have at hand, but we will be
able to say something explicit only for the case $\l=1/2$ and  
$\m =\n =\kappa =0$, which is related to the 
$SO(2) \times SO(2) \times SO(2)$ model of five-dimensional gauged
supergravity or to the $SO(2) \times SO(2) \times SO(2) \times SO(2)$ 
model in four dimensions.        

We proceed by considering the Lam\'e equation 
\be
\left( -{d^2 \over du^2} + n(n+1) \wp(u)\right)\Psi(u) 
= E \Psi(u) ~,
\ee
where $E$ are the energy levels of the corresponding one-dimensional
quantum mechanical problem and $\Psi(u)$ are normalized wave functions
that vanish (typically) at $u=0$ modulo the real period $2\omega_1$. 
Of course, when $n$ is a positive integer, the solutions are easily
described using appropriate ratios of the Weierstrass sigma-function, 
namely
\ba 
\Psi_1(u) &=& e^{\a u}{\sigma(u-\b_1) \sigma(u-\b_2)  
\cdots \sigma(u-\b_n) \over
\sigma^n(u)} ~, \nonumber\\ 
\Psi_2(u) &=& e^{-\a u}{\sigma(u+\b_1) \sigma(u+\b_2)  
\cdots \sigma(u+\b_n) \over
\sigma^n(u)} ~,  
\ea
where the constants $\a$ and $\b_i$ are determined by substituting the
ansatz into Lam\'e's equation. Using these two (in general independent)
solutions one can construct regular solutions by taking suitable
linear combinations of them and obtain a transcendental equation for the 
energy eigenvalues. 

On the other hand, if $n$ is not an integer the
solutions will be difficult to describe, even in a formal sense.    
Some simplifications occur when $n$ is half of an odd positive integer.
It is known in this case, using the substitution
\be
\Psi = \left(\wp^{\prime}(z)\right)^{-n} \Phi (z) ~, ~~~~~
{\rm where} ~~ z=u/2  \ ,
\ee
that Lam\'e's equation transforms into the differential equation
\be
{d^2 \Phi \over dz^2} -2n {\wp^{\prime \prime}(z) \over 
\wp^{\prime}(z)} {d \Phi \over dz} + 4\left(n(2n-1) \wp(z)
+ E\right) \Phi = 0 ~. 
\ee
Then, according to results obtained by Brioschi and Halphen more than
a century ago, 
a formal solution can be written as
\be
\Phi(z) = \sum_{r=0}^{\infty} c_r \left(\wp(z) - e_2 \right)^{a-r} ~,
\ee
provided that 
\be
(a-2n)(a-n+1/2) = 0 \ ,
\ee
and that the following recursive relations are satisfied:

\vskip .2cm

$(a-r-2n)(a-r-n+1/2)c_r + \left(3e_2(a-r+1)(a-r-2n+1) + e_2 n(2n-1) 
+ E \right)c_{r-1}$ 

\vskip -.2cm 

\be
= (e_1 -e_2)(e_2 -e_3)(a-r+2)(a-r-n+3/2) c_{r-2} ~. 
\ee
Here, $e_1$, $e_2$ and $e_3$ denote the three roots of the cubic
curve in its Weierstrass form. 
Actually, when $n$ is half of an odd positive integer, 
there is a solution expressible in finite form 
\be
\Phi(z) = \sum_{r=0}^{n-1/2} c_r \left(\wp(z) -e_2 
\right)^{2n - r} ~,  
\ee
which corresponds to $a=2n$, and provides discrete energy levels $E$
by solving the recursive relations with $c_{n+1/2} = 0$; they are all
real for curves having real roots $e_i$. Otherwise, for solutions
expressible as an infinite sum, the energy levels remain arbitrary by
these general considerations alone. 

The simplest case to consider has $n=1/2$. It turns out that a solution
expressible in finite form has only one term:
\be
\Phi(z) = c_0 \left(\wp(z) - e_2 \right) \ ,
\ee
with energy $E=0$ and $c_0$ arbitrary. This yields one of the two
independent solutions of Lam\'e's equation with zero energy, call it
$\Psi_1(u)$, whereas the other is obtained by employing the general
formula
that relates any two solutions with the same energy $E$: 
\be
\Psi_2(u) = C \Psi_1 (u) \int {du \over {\Psi_1}^2(u)} ~,  
\ee
where $C$ is another constant. An explicit calculation shows that the 
most general solution with $E=0$ is given in terms of two integration
constants $A$, $B$ by
\be
\Psi(u) = {1 \over \sqrt{\wp^{\prime} (u/2)}} \left(A \wp(u/2) 
+ B \right) ~. 
\ee
Having established this, one may impose the regularity of the physical
solution at $u=0$, namely $\Psi(0) = 0$, and set the coefficient
$A = 0$; otherwise the wave function will diverge as 
$1/\sqrt{u}$ for
$u \rightarrow 0$. For $u = 2\omega_1$, however, the wave-function blows up
and hence it is not normalized.  
 
For $n=1/2$, but with $E \neq 0$, the formal solution has an infinite
number of terms. According to the general discussion, we find after
some calculation (working with either $a=0$ or $a=1$) the result
\be
\Phi(z) = C \left({1 \over E} + \sum_{k=1}^{\infty} 
{d_k \over \left( \wp(z) - e_2\right)^k} \right) ~,  
\ee
where $C$ is an arbitrary constant and 
\ba
d_1 & = & -{1 \over 2} ~, ~~~~~ d_2 = {1 \over 2}\left({E \over 6} 
+ e_2 \right) ~, \nonumber\\
d_3 & = & -{1 \over 12} \left((e_1 - e_2)(e_2 - e_3) + 9
\left({E \over 6} + e_2 \right) \left({E \over 18} + e_2 \right) \right)
~, \\
d_4 & = & {3 \over 20} \left((e_1 - e_2)(e_2 - e_3) 
\left({7E \over 36} + 2 e_2 \right) + 9 \left({E \over 6} + e_2 \right)
\left({E \over 18} + e_2 \right) \left( {E \over 36} + e_2 \right) 
\right) \ , \nonumber
\ea
and so on. Unfortunately, even in this simplest case with $n = 1/2$, 
it is very difficult 
to solve the recursive relations and find all coefficients
$d_k$ in closed form in order to sum up the infinite series of terms.
In any case, this procedure yields one formal solution of Lam\'e's equation
with $E \neq 0$, $\Psi_1 (u)$, whereas the other (independent) 
formal solution 
of the same energy, $\Psi_2(u)$, 
can be obtained according to the general formula above.
However, the integration that determines $\Psi_2(u)$ in terms of 
$\Psi_1(u)$ cannot be performed explicitly unless one knows first how to
sum up the infinite series of terms for $\Phi(z)$. 

Note that for $n=1/2$ the
formal solution $\Psi_1(u)$ appears to be regular at $u=0$, whereas at 
$u = 2 \omega_1$ (where $ \wp(u/2) = e_1$) this is not guaranteed.
In fact, since the coefficients $d_k$ are polynomial functions of the 
energy $E$, one must demand that the resulting ``energy series" converge.
Since $\wp^{\prime} (\omega_1) =0$, we demand 
i.e.
\be
{1\ov E}+
\sum_{k=1}^{\infty} {d_k (E) \over (e_1 - e_2)^k} =0\ .
\ee
This is certainly a non-trivial constraint on the allowed energy bands
when $E \neq 0$, which also depend on the relative size of the $a$- and
$b$-cycles of the Riemann surface, i.e. the differences 
$e_2 - e_3$ and $e_2 - e_1$ that appear in the ``energy sum". 
However, we are not able at present to find the complete
solution to the problem in closed (even transcendental) form.

\section{Conclusions}

We have investigated in detail the structure of domain wall
solutions in theories of gauged supergravity in diverse
dimensions by considering the effect of non-trivial scalar
fields taking values in the coset space $SL(N, \IR)/SO(N)$.
The presentation was kept quite general to cover, where possible,
aspects of domain wall solutions in theories of D-dimensional
gravity by turning off the effect of any other fields, such
as gauge fields and fermions. Special emphasis has been placed
on two cases, namely $(D, N) = (4, 8)$ and $(7, 5)$, which arise
by compactification of eleven-dimensional supergravity on 
$S^7$ and $S^4$ respectively. 
The effect of
the scalar fields in four (or seven) dimensions is related to 
deformations of the round spheres in the compactifying space,
thus breaking the isometry group $SO(8)$ (or $SO(5)$) into
appropriate symmetry subgroups. In fact, we were able to give an
algebraic classification of all such cases using the 
Christoffel--Schwarz transformation, which arises in the solution
of the first-order Bogomol'nyi-type equations for the conformal
factor and the scalar fields of these models. As a result, for
$D=4$, we found a hierarchy of 22 solutions 
starting from an algebraic curve of genus 9 
corresponding to completely broken isometry group.
When some cycles shrink to zero size, by letting some moduli
coalesce, the symmetry group is enhanced, whereas the genus
of the Riemann surface is lowered accordingly. Among the genus
0 models there is the $AdS_4$ space with no scalar fields, having 
maximal isometry group $SO(8)$, while in all other cases $AdS_4$
is reached only asymptotically. Similarly, for $D=7$, we found
7 models in total, which can be classified starting from a genus
6 algebraic curve with no isometry and proceeding all the way 
down to genus 0, where the $AdS_7$ space arises with maximal
isometry group $SO(5)$. All other models in the list admit subgroups
of $SO(5)$ as isometries and approach $AdS_7$ only asymptotically
due to the presence of non-trivial scalar fields.

A geometrical picture of our solutions in four and seven
dimensions is provided in terms of distributions of M2- and M5-branes
in eleven dimensions. 
The analysis has been carried out in detail
choosing suitable harmonic functions that describe continuous 
distributions of branes in eleven dimensions, and is in accordance
with the geometric deformation of seven and four dimensional
spheres induced by the non-trivial moduli of the underlying
algebraic curves. The resulting picture resembles the construction
of domain walls of five-dimensional gauged supergravity with 
non-trivial scalar fields in the coset $SL(6, \IR)/SO(6)$, where
suitable continuous distributions of D3-branes were considered
in ten-dimensional type-IIB supergravity. An interesting problem 
that remains open for further study is the possible effect of dualities on
the structure of domain-wall (and other) solutions in various dimensions. 
For example, M2- and M5-branes are related to each other via 
electric-magnetic duality, and hence various distributions of them
in eleven dimensions should yield strong-weak coupling relations
among solutions of gauged supergravities in lower dimensions.
Also, D3-branes appear in a sequence of dualities between 
extended objects in higher dimensions, using both S- and T-dualities,
and so the variety of domain walls in $D= 4, ~ 5$ and 7 dimensions (where
consistent truncations of supergravity are known to exist) ought
to be interelated. Of course, it will be interesting to formulate
this in the algebro-geometric context provided by the 
Christoffel--Schwarz transformation, where there is a universal moduli
space of the genus 9 Riemann surface for M2-branes  
$y^4 = (x-b_1)(x-b_2) \cdots (x-b_8)$. It has been noted earlier that
the Riemann surfaces of D3-branes arise as special cases 
by taking two of the moduli to
infinity, while for M5-branes there are three moduli taken to infinity.
Thus, it is quite natural to expect that the modular transformations
of the underlying algebraic curves have a natural interpretation in 
higher dimensions as S- and T-dualities among branes and their 
continuous distributions thereof. We hope to return to this issue in
a separate publication. 

Another topic that has been investigated in detail concerns the
analytic form of fluctuations for the graviton and scalar
fields in our domain-wall backgrounds. Since the quantitative 
analysis of this problem amounts to solving a one-dimensional
quantum mechanical problem with potential $V(z)$ that behaves as
$1/z^2$ for $z \rightarrow 0$ (the asymptotic AdS region), the 
right identification of the variable $z$ and the associated 
Schr\"odinger potential $V(z)$ become crucial for extracting
the spectrum. We found that the variable $z$ appears naturally in
the Christoffel--Schwarz transformation, which is a complex
transformation that maps the interior of a closed polygon in 
the $z$-plane onto the upper-half $F$-plane, while the variable
$F$ is more appropriate for the brane description of our configurations
in higher dimensions. Thus, the uniformization of the associated
algebraic curves is a necessary step 
in order to derive the exact form of the potential, 
in each case of interest, and, consequently, to determine 
its spectrum. We were able to complete this 
mathematical task for the models with Riemann surfaces of genus 0 
and 1 and derive the potentials in closed form for most of these 
cases. Subsequently, using techniques of supersymmetric quantum
mechanics, which ensure that the spectrum is non-negative, we 
were able to compute the spectrum exactly for many models and
make estimates using the WKB approximation for many others. The
elliptic models exhibit an interesting class of potentials within
the family of generalized Lam\'e potentials, which can be studied
analytically only in certain cases. 

It might seem surprising that we have encountered domain wall 
solutions with very little or even no isometry in the classification
in terms of Riemann surfaces. Of course, we are unable to
perform the uniformization and find explicit expressions for the
conformal factor of their metrics and their scalar fields 
in general, since this 
has only been done for models of low genus and hence bigger 
isometry groups. Nevertheless, this situation can be perfectly 
accommodated in supersymmetric theories, where all the domain-wall
solutions leave half of the supersymmetries unbroken. 
It is rather
instructive to compare this situation with the geometry of 
four-dimensional hyper-K\"ahler manifolds and their possible isometry
groups. 
Recall that all hyper-K\"ahler manifolds are supersymmetric (preserving
half of the supersymmetries) and as such they admit three independent
complex structures $I$, $J$, $K$, hence a whole sphere of them, 
since $aI + bJ + cK$ will be a complex structure if $a^2 +b^2 +c^2 =1$.
The group $SO(3)$ acts naturally on the space of complex structures 
as rotations; 
but this does not necessarily imply that all 
hyper-K\"ahler manifolds are $SO(3)$ symmetric. Although many examples
(like Eguchi--Hanson, Taub--NUT and Atiyah--Hitchin) have an $SO(3)$
isometry group, there are others with less isometry or none; for example
Dancer's manifold has only an $SO(2)$ isometry, whereas K3 commonly
used in compactifications of string theory has no isometries at all.
In this sense, our domain-wall solutions to gauged supergravities in 
four, five and seven dimensions provide useful tools for 
developing a deeper understanding of the consistent truncations 
of eleven-dimensional supergravity, and they can be rather exotic.  
Also, the associated Bogomol'nyi bounds and their possible description
using contour integrals on the underlying Riemann surfaces pose some
interesting  mathematical questions for the future.

Finally, there is the conceptual 
issue of relating ${\cal R}$-symmetry to the isometry group that 
remains unbroken by the geometric structure of our solutions, which
we belief is worth emphasizing. In our approach, the ${\cal R}$-symmetry
is spontaneously broken by giving vacuum expectation values to 
the scalar fields 
of the theory that are charged under this symmetry. 
Then, the ${\cal R}$-symmetry is not a symmetry
anymore, but relates different vacua of the theory. Generically, this 
procedure also 
breaks conformal invariance, thus breaking half of the 32 supersymmetries 
independently of the amount of symmetry that is left unbroken. 
In supergravity this can be understood as follows: for generic
values of the real parameters $b_i$ in the master curve
$y^4 = (x-b_1)(x-b_2) \cdots (x-b_8)$,  the gauge symmetry
(related to the ${\cal R}$-symmetry in field theory), is spontaneously
broken. In fact, for any given choice of the parameters, the curve
is not invariant under $SO(8)$. However, its form is preserved since
$SO(8)$ acts naturally on $b_1, ~b_2, \cdots , b_8$  in its fundamental
representation and rotates any given choice of moduli $b_i$ into another.
This is different from the situtation where the background describes a flow
from the maximally supersymmetric theory to a conformal theory with less
supersymmetry. In the latter case the theory is perturbed by adding 
suitable deformations
to the action, which explicitly break some or all of supersymmetries;
at the IR fixed point the geometry is again $AdS_D$ and the 
${\cal R}$-symmetry is related to the number of supersymmetries. This
means, in particular, 
that the isometry group contains a factor that is equal to the
${\cal R}$-symmetry of the field theory.

\end{document}